\documentclass[%
superscriptaddress,
reprint,
amsmath,amssymb,
aps,
pre,
]{revtex4-2}

\usepackage[english]{babel}

\usepackage{graphicx}
\usepackage[percent]{overpic}
\usepackage{dcolumn}
\usepackage{bm}

\usepackage{hyperref}
\usepackage{color}
\definecolor{rltred}{rgb}{0.75,0,0}
\definecolor{rltgreen}{rgb}{0,0.6,0}
\definecolor{rltblue}{rgb}{0.3,0.3,1}

\usepackage{placeins}

\hypersetup{colorlinks,%
	hypertexnames=true,%
	linkcolor=rltred,%
	citecolor=rltgreen,%
	linktocpage=true}

\usepackage[english]{babel}

\begin{document}
\title{Reduced density matrix description of high-harmonic generation in multi-electron atoms: exploring sub-cycle correlation effects}

\author{Katharina Buczolich}
\email{katharina.buczolich@tuwien.ac.at}
\affiliation{Institute for Theoretical Physics, Vienna University of Technology,
	Wiedner Hauptstra\ss e 8-10/136, 1040 Vienna, Austria, EU}
 
\author{Takeshi Sato}
\affiliation{Graduate School of Engineering and Photon
Science Center, School of Engineering, The University of Tokyo, Bunkyo-ku, Tokyo 113-8656, Japan, \\
Research Institute for Photon Science and Laser Technology, The University of Tokyo, Bunkyo-ku, Tokyo 113-0033, Japan, \\
and Department of Nuclear Engineering and Management, Graduate School of Engineering, The University of Tokyo,7-3-1 Hongo, Bunkyo-ku, Tokyo 113-8656, Japan}

\author{Kenichi L. Ishikawa}
\affiliation{Graduate School of Engineering and Photon
Science Center, School of Engineering, The University of Tokyo, Bunkyo-ku, Tokyo 113-8656, Japan, \\
Research Institute for Photon Science and Laser Technology, The University of Tokyo, Bunkyo-ku, Tokyo 113-0033, Japan, \\
and Department of Nuclear Engineering and Management, Graduate School of Engineering, The University of Tokyo,7-3-1 Hongo, Bunkyo-ku, Tokyo 113-8656, Japan}

\author{Fabian Lackner}
\affiliation{Institute for Theoretical Physics, Vienna University of Technology, Wiedner Hauptstra\ss e 8-10/136, 1040 Vienna, Austria, EU}

\author{Joachim Burgd\"orfer}
\affiliation{Institute for Theoretical Physics, Vienna University of Technology, Wiedner Hauptstra\ss e 8-10/136, 1040 Vienna, Austria, EU}

\author{Iva B\v rezinov\'a}
\email{iva.brezinova@tuwien.ac.at}
\affiliation{Institute for Theoretical Physics, Vienna University of Technology,
	Wiedner Hauptstra\ss e 8-10/136, 1040 Vienna, Austria, EU}

\date{\today}

\begin{abstract}
High-harmonic generation (HHG) is one of the fundamental processes at the heart of attosecond physics. Traditionally viewed as an effective single-particle effect, recent advances have focused on contributions beyond this single-particle picture to the harmonic spectrum, as well as on probing electron correlations through HHG in atoms, molecules, and solids. In this paper, we introduce a reduced density matrix description to explore and to quantify correlation effects on a sub-cycle time scale and apply this approach to prototypical multi-electron atoms. By comparing noble gas atoms (He, Ne) with alkaline earth atoms (Be, Mg) exposed to driving fields with similar Keldysh parameters, we show that 
the sub-cycle variation of correlation parameters differs markedly for noble gases and alkaline earth atoms. We provide an intuitive explanation of these surprising effects based on the dynamics of natural orbitals and discuss the effect of this ultrafast correlation dynamics on the HHG spectrum.
\end{abstract}

\maketitle

\section{Introduction}\label{sec:intro}
One of the most striking signatures of the non-linear responses of matter to driving with strong fields is the generation of high harmonics. First observed in atomic vapors irradiated by strong infrared (IR) fields \cite{mcpherson_studies_1987, ferray_multiple-harmonic_1988}, high-harmonic generation (HHG) has developed into a powerful tool for generating attosecond pulse trains \cite{schins_observation_1994, veniard_phase_1996, paul_observation_2001, aseyev_attosecond_2003, lopez-martens_amplitude_2005}, isolated attosecond pulses \cite{nisoli_compression_1997, drescher_time-resolved_2002, kienberger_atomic_2004, santra_theory_2011, chen_light-induced_2012}, and tunable tabletop sources of extreme ultraviolet and X-ray radiation \cite{spielmann_generation_1997, popmintchev_bright_2012, hadrich_exploring_2015, naranjo-montoya_table-top_2024}, giving rise to the new field of attosecond physics \cite{brabec_intense_2000, krausz_attosecond_2009}. The appearance of the HHG spectrum extending into the XUV regime was readily identified as a clear signature of the breakdown of multi-photon perturbation theory and explained by the so-called three-step model \cite{Kuchiev.45.319, corkum_plasma_1993, lewenstein_theory_1994}. It provides an intuitive (semi-classical) picture of HHG in terms of a sequence of three subsequent one-electron processes: tunneling (or over-the-barrier) ionization near the peak of the strong IR field lowering the potential barrier, followed by acceleration of the ionized electron along a trajectory that will eventually steer the electron back to the ion with high kinetic energy, and finally the radiative recombination with the parent ion converting the acquired kinetic energy into high-energy radiation. This single-active electron (SAE) picture is at the core of several variants of the so-called strong-field approximation (SFA) as well as of many numerical simulations, capturing in many cases the essential features of the HHG spectrum.\\
More recently, several features have been observed that cannot be explained within the single-electron framework because of the crucial role of multi-electron effects. One prominent example is the giant dipole resonance in Xe \cite{shiner_probing_2011, pabst_strong-field_2013}, where the interplay between the $5\text{p}$ and $4\text{d}$ shells upon recombination leads to a dramatic enhancement in the yield near $100$ $\text{eV}$ \cite{shiner_probing_2011,pabst_strong-field_2013}. Enhancements typically spectrally limited to one or a few harmonics have been observed (see e.g.~\cite{ganeev_strong_2006,ganeev_strong_2007,ganeev_experimental_2012,haessler_phase_2013}) and theoretically analyzed in other multi-electron atoms and ions (see e.g.~\cite{wahyutama_time-dependent_2019, bray_correlation-enhanced_2020}). Another prominent example, where multi-electron correlations have dramatic effects on the yield in HHG, is the secondary plateau that extends far beyond the energy cutoff expected for the most weakly bound electron. The emergence of secondary plateaus has been observed numerically e.g.~in He \cite{artemyev_impact_2017} and has been traced back to the correlated electron dynamics. The mechanism identified for He \cite{de_las_heras_spectral_2020} corresponds to the correlated back action of one electron on the other during tunnel ionization which leaves the ion in an excited state such that the electron returning to the ion sees a larger ionization potential. In turn, in Be ionized by strong laser fields, the appearance of the extended plateau was attributed to a recollision-induced decrease in the two-electron ionization probability such that recombination occurs into the $\text{Be}^{2+}$ ion \cite{tikhomirov_high-harmonic_2017}. A similar mechanism was later found to lead to extended plateaus also in Li and Na \cite{li_high-order_2019}. A systematic numerical study of noble gas atoms within time-dependent density functional theory (TDDFT) showed that correlations strongly influence the harmonic yield as a function of the ellipticity of the driving field \cite{neufeld_probing_2020}.\\
By applying HHG spectroscopy to extended systems and solids \cite{ghimire_observation_2011, ghimire_strong-field_2014, luu_extreme_2015, vampa_all-optical_2015, floss_ab_2018} it was numerically demonstrated that strong driving can lead to the ultrafast formation of correlated pair-states called doublons for the strongly correlated Fermi-Hubbard system whose dynamics leaves marked finger prints on the HHG spectrum \cite{silva_high-harmonic_2018}. This observation suggests HHG as a potential spectroscopic tool to investigate the ultrafast dynamics of strongly correlated materials. For example, theoretical investigations of HHG in strongly correlated materials have pointed to dynamical modifications of effective interaction parameters \cite{tancogne-dejean_ultrafast_2018}. Such a dynamical renormalization of the Coulomb repulsion has been, indeed, subsequently measured in a strongly-correlated cuprate material upon strong-field driving \cite{baykusheva_ultrafast_2022}.\\
In this paper, we describe and quantify the role of electronic correlations by introducing a reduced density matrix (RDM) description for HHG that allows to express the radiation amplitude according to the Larmor formula in terms of a superposition of independent radiating oscillators for any correlated many-electron system. This description extends the very first theoretical approach to HHG by Kuchiev \cite{Kuchiev.45.319} in terms of ``atomic antennas" to correlated systems. We show that these antennas can be fully characterized by the natural orbitals and their occupation numbers of the many-electron system. Employing correlation and entanglement measures from quantum information theory, ultrafast variations of electron correlations on sub-cycle time scales can be extracted from the time-dependent occupation numbers of these natural orbitals. We apply this approach to prototypical multi-electron noble gas (He, Ne) and alkaline earth (Be, Mg) atoms based on fully converged multi-configuration time-dependent Hartree-Fock (MCTDHF) \cite{caillat_correlated_2005, kato_time-dependent_2004, nest_multiconfiguration_2005, haxton_multiconfiguration_2011, karabanov_accuracy_2011, sato_time-dependent_2016} calculations. We show that strongly driven noble gas and alkaline earth atoms feature markedly different sub-cycle dynamics of the time-dependent correlations, referred to in the following also as dynamical correlations, and discuss their implications for the resulting HHG spectra. \\
The paper is structured as follows. In Sec.~\ref{sec:rdms} we derive within the framework of the Larmor formula of classical radiation theory an explicit expression for the amplitude of HHG radiation in terms of the one-particle reduced density matrix (1RDM) for an arbitrary $N$-electron system. The equations of motion for the 1RDM under strong-field driving by the laser field will be analyzed in Sec.~\ref{sec:nat_orb}. Measures for time-dependent correlations to quantify the correlation dynamics will be introduced in Sec.~\ref{sec:corr_meas}. Numerical applications to noble gas and alkaline earth atoms will be presented in Sec.~\ref{sec:hhg_corr}. Conclusions and an outlook to future extensions of this work will be given in Sec.~\ref{sec:conclusions}. Unless stated otherwise, we use atomic units throughout the paper, i.e.~ $\hbar = m_e = e = 1$.
\section{Reduced density matrix representation of HHG}\label{sec:rdms}
We consider in the following an interacting $N$-particle system driven by a strong laser field and governed by the time-dependent Schr\"odinger equation
\begin{equation}\label{eq:tdse}
    i\partial_t |\Psi(t)\rangle = \hat H(t) |\Psi(t)\rangle.
\end{equation}
The general form of the Hamiltonian $\hat H(t)$ is given by
\begin{equation}\label{eq:hamil}
    \hat H(t) =  \sum_{n=1}^N \hat h_n(t)+\hat V_n(t) + \frac{1}{2}\sum_{n\neq m}^N \hat W_{nm},
\end{equation}
where $\hat h_n(t)$ stands for the single-particle operators of the unperturbed system, which in our application to atoms will include the kinetic energy operator and the nuclear Coulomb potential. The coupling to the strong laser field is given by the single-particle operator $\hat V_n(t)$ which takes the form $\hat{V}_n^{(L)}(t)=\hat z_n F(t) $ in the length gauge and $\hat V_n^{(V)}(t)=\hat p_{z_n}A(t)/c$ in the velocity gauge with $A(t)$ the vector potential associated with the strong electric field $F(t)$ oriented in the following along the $z$-axis. For atomic systems, the electron-electron Coulomb interaction will play the role of the inter-particle interaction $\hat W_{nm}$. It should be noted that Eq.~\ref{eq:hamil} and the following description is general and applies also to systems featuring short-ranged interactions and to extended systems as well.\\
Within the classical description of the emitted radiation field underlying the Larmor formula, the high-harmonic yield is given by
\begin{equation}\label{eq:Larmor}
    I^\text{HHG} = \frac{2}{3c}\left |\tilde{\bm{{a}}}(\omega)\right|^2
\end{equation}
with $\tilde{\bm{a}}(\omega)$ being the Fourier transform of the dipole acceleration 
\begin{align}\label{eq:dipole_a}
    \bm{a}(t) &= -\frac{d^2}{dt^2} \sum_{n=1}^N\langle \Psi(t)|\hat{\bm{r}}_n|\Psi(t)\rangle \nonumber \\
    &=  \frac{d}{dt} \sum_{n=1}^N\langle \Psi(t)|\hat{\bm{p}}_n|\Psi(t)\rangle \nonumber \\
    &= i\sum_{n=1}^N\langle \Psi(t)|[\hat h_n(t),{\hat{\bm{p}}}_n]|\Psi(t)\rangle \nonumber \\
    &= \sum_{n=1}^ N \langle \Psi(t)|\hat{\bm{a}}_n|\Psi(t)\rangle.
\end{align}
The operator of the dipole acceleration $\hat{\bm{a}}_n=i[\hat h_n(t),\hat{\bm{p}}_n]$ is proportional to the spatial gradient of the one-particle potential terms in Eq.~\ref{eq:hamil}, i.e.~the nuclear Coulomb potential in the present case. With the use of Eq.~\ref{eq:Larmor}, quantum effects of the photonic field are neglected from the outset.\\
Since the physical observable of interest, the acceleration $\hat{\bm{a}} = \sum_{n=1}^ N \hat{\bm{a}}_n$, is a one-body operator, its expectation value (Eq.~\ref{eq:dipole_a}) evaluated for the exact time-dependent many-electron state $|\Psi(t)\rangle$ (Eq.~\ref{eq:tdse}) can be equivalently expressed by the one-particle reduced density matrix (1RDM) $D_1$ without invoking any approximation in the electronic degrees of freedom. The 1RDM is obtained from the many-body wavefunction $|\Psi(t)\rangle$ by tracing out all but one particle
\begin{equation}\label{eq:1rdm}
    D_1(t) = N \text{Tr}_{2\hdots N}|\Psi(t)\rangle\langle \Psi(t)|.
\end{equation}
We label here and in the following all one-particle operators by the subscript 1. In terms of $D_1$ (Eq.~\ref{eq:1rdm}), Eq.~\ref{eq:dipole_a} reduces to 
\begin{align}\label{eq:dipole_a_1rdm}
    \bm{a}(t) &= \text{Tr}\left[{\hat{\bm{a}}}_1(t)D_1(t)\right] .
\end{align}
At any instant of time, the 1RDM can be diagonalized in terms of the time-dependent natural orbitals $\{|\eta_j(t)\rangle \}$  and their eigenvalues $\eta_j(t)$, the occupation numbers of these orbitals $| \eta_j(t) \rangle$,
\begin{equation}\label{eq:1rdm_eig}
    D_1(t) = \sum_j \eta_j(t) |\eta_j(t)\rangle \langle \eta_j(t)|,
\end{equation}
with $j$ being a combined index of a spatial orbital and the spin ($j = q_j, \sigma_j$), $|\eta_j\rangle  = |\nu_{q_j}\rangle \otimes |\sigma_j\rangle $. They play a key role in the analysis of time-dependent, i.e.~dynamical correlations. Within any mean-field approach such as the time-dependent Hartree-Fock approximation (TDHF) \cite{kulander_schafer_krause_1992}, the TDDFT \cite{runge_density-functional_1984} or any SAE approximation, Eq.~\ref{eq:1rdm_eig} reduces to 
\begin{align}\label{eq:1rdm_mean_field}
    D_1(t)\approx \sum_{j=1}^ {N_\text{occ}} |\eta_j(t)\rangle \langle \eta_j(t)|,
\end{align}
i.e.~the occupation numbers of such mean-field (spin) orbitals are 
\begin{align}\label{eq:MF_nat_occ}
 \eta_j(t) &=
  \begin{cases}
   1        & j\leq N_\text{occ}\\
   0        & j> N_\text{occ},
  \end{cases}
\end{align}
where $N_\text{occ}$ is the number of occupied orbitals. One crucial difference that distinguishes the fully correlated $N$-electron dynamics (Eq.~\ref{eq:1rdm_eig}) from the mean-field description (Eq.~\ref{eq:1rdm_mean_field}) is that, in general, a larger number of fractionally occupied natural orbitals contribute, the number of which being decoupled from the number of electrons $N$ in the system. The occupation numbers are only constrained by 
\begin{equation} \label{eq:sum_occ}
    \sum_j \eta_j(t) =N.
\end{equation}
Consequently, the expression for the dipole acceleration (Eq.~\ref{eq:dipole_a}) reads 
\begin{equation}\label{eq:dipole_a_nat}
     \bm{a}(t) = \sum_{j} \eta_j(t) \langle \eta_j(t)|\hat{\bm a}_1|\eta_j(t)\rangle .
\end{equation}
Thus, the amplitude of the emitted high-harmonic radiation is given by the superposition of the fields generated by an ensemble of strongly driven charge oscillators or, in the terminology of Kuchiev's pioneering early work on HHG \cite{Kuchiev.45.319}, by an ensemble of ``atomic antennas". The effective oscillator strength of these antennas is controlled by the time-dependent dipole acceleration expectation values $\langle\eta_j(t)|\hat{\bm a}_1|\eta_j(t) \rangle$ and the occupation numbers, $\eta_j(t)$. The strength of these oscillations introduced in Eqs.~\ref{eq:sum_occ} and \ref{eq:dipole_a_nat} should not be confused with the conventional oscillator strength for atomic dipole transitions in the perturbative limit. The number of such ``atomic antennas" ($\geq N$) that effectively contribute is controlled by the strength of the static and dynamical correlations present in the driven system.\\
We note that Eq.~\ref{eq:dipole_a_nat} is general and equally applicable to atoms, molecules, and solids, and is independent of the strength of correlations. Apart from neglecting effects of the quantized radiation field by way of using the Larmor formula, Eq.~\ref{eq:dipole_a_nat} is exact as far as the electronic degrees of freedom are concerned. Moreover, Eq.~\ref{eq:dipole_a_nat} can be generalized to other physical observables that can be written as sums of expectation values of one-body operators. In the specific case of HHG in multi-electron atoms, the expectation value of the dipole acceleration initially vanishes prior to the action of the symmetry-breaking driving field but will become non-zero as a function of the time when the laser field starts to polarize the state creating a state of mixed parity, i.e.~containing angular momentum contributions of both even and odd $l$. Furthermore, Eq.~\ref{eq:dipole_a_nat} suggests that the intuitive picture of the three-step model for HHG should be mirrored by and can be mapped onto the time evolution of each of the individual ``antennas", their expectation values and occupation numbers, even when the independent-particle model fails and electron correlations significantly contribute.
\section{Time evolution of natural orbitals}\label{sec:nat_orb}
The time dependence of the dipole acceleration (Eq.~\ref{eq:dipole_a_nat}) which governs the resulting HHG spectrum (Eq.~\ref{eq:Larmor}) is completely determined by the time-dependence of the natural orbitals $|\eta_j(t)\rangle$ and of their eigenvalues $\eta_j(t)$. It is therefore of considerable conceptual interest to explicitly determine the equations of motion for $|\eta_j(t)\rangle$ and $\eta_j(t)$ for the strongly driven $N$-electron systems.\\
Starting point is the equation of motion of the 1RDM which is the lowest-order member of the Bogoliubov-Born-Green-Kirkwood-Yvon hierarchy \cite{bogoljubov_lectures_1985} and takes the form of a Liouville-von Neumann equation including a collision term that couples the 1RDM to the two-particle reduced density matrix (2RDM) $D_{12}$,
\begin{equation}\label{eq:eom_D1}
    i\partial_t D_1 = [\hat h_1, D_1] + \text{Tr}_2[\hat W_{12},D_{12}].
\end{equation}
The 2RDM is given by 
\begin{equation}\label{eq:2rdm}
    D_{12}(t) = N(N-1)\text{Tr}_{3\hdots N}|\Psi(r)\rangle\langle \Psi(t)|.
\end{equation}
The collision term (second term on the rhs of Eq.~\ref{eq:eom_D1}) represents the interaction of particle $1$ with all other particles surrounding it. This term includes both mean-field contributions as well as electron-electron (or, more generally, pair-)correlations. In order to separate these two contributions from each other, we employ the decomposition of the 2RDM into an uncorrelated (or Hartree-Fock) term and the two-particle correlation term as
\begin{equation}\label{eq:cum_expansion}
    D_{12}(t) = \hat A D_1(t)D_2(t) + \Delta_{12}(t).
\end{equation}
In Eq.~\ref{eq:cum_expansion}, $\hat A$ denotes the anti-symmetrization operator that anti-symmetrizes the product of the two 1RDMs, and $\Delta_{12}$ denotes the two-particle correlator frequently also referred to as the two-particle cumulant \cite{mazziotti_approximate_1998, kutzelnigg_cumulant_1999}. Projecting onto the basis of natural orbitals, i.e.~the eigenstates of the 1RDM at each instant of time, the matrix representation of Eq.~\ref{eq:eom_D1} becomes (for reasons of brevity we omit the explicit time-dependent of $\eta_j(t)$ and $|\eta_j(t)\rangle$ in the notation of Eq.~\ref{eq:1rdm_eom_nat}-\ref{eq:orthonorm_dot} below)
\begin{align}\label{eq:1rdm_eom_nat}
    \langle \eta_i | i \partial_t D_1 | \eta_j \rangle = \langle \eta_i | [\hat{h}_1+\hat{V}_1(t), D_1] + \text{Tr}_2[\hat{W}_{12}, D_{12}] | \eta_j \rangle  \nonumber \\
    = ( \eta_j  - \eta_i ) \langle \eta_i | \hat{h}_1 +\hat V_1(t)| \eta_j \rangle + \langle \eta_i| \text{Tr}_2 [\hat{W}_{12}, D_{12}] | \eta_j \rangle.
\end{align}
On the other hand, in the representation of the 1RDM in the natural-orbital basis (Eq.~\ref{eq:1rdm_eig}) the equation of motion can be equivalently written as 
\begin{equation}\label{eq:1rdm_dots}
  \langle \eta_i | i \partial_t D_1 | \eta_j \rangle = i \dot{\eta}_i \langle \eta_i | \eta_j \rangle + i \eta_j \langle \eta_i | \dot{\eta}_j \rangle + i \eta_i \langle \dot{\eta}_i | \eta_j \rangle.
\end{equation}
Using $\langle \eta_i|\eta_j\rangle=\delta_{ij}$ and 
\begin{equation}\label{eq:orthonorm_dot}
    \frac{d}{dt}\langle \eta_i|\eta_j\rangle =  \langle \eta_i | \dot{\eta}_j \rangle + \langle \dot{\eta}_i | \eta_j \rangle = 0,
\end{equation}
equating Eqs.~\ref{eq:1rdm_eom_nat} and \ref{eq:1rdm_dots}, and inserting Eq.~\ref{eq:cum_expansion} yields for the diagonal elements $i=j$ the equation of motion for the natural orbital occupation numbers 
\begin{align}\label{eq:eom_nat_occ}
    i\partial_t \eta_j(t) &= \langle \eta_j(t)|\text{Tr}_2[\hat W_{12},\Delta_{12}(t)|\eta_j(t)\rangle \nonumber \\
    &=\langle \eta_j(t)|\hat{C}_1(t)|\eta_j(t)\rangle.
\end{align}
In Eq.~\ref{eq:eom_nat_occ}, we have introduced the reduced collision operator $\hat{C}_1$ whose spatial and spin representation reads 
\begin{align}
    & C_1(\bm r_1, \sigma_1, \bm r_1', \sigma_1'; t)= \nonumber \\
    & \sum_{\sigma_2} \int d^3r_2\left[W_{12}(\bm r_1-\bm r_2)-W_{12}(\bm r'_1-\bm r_2)\right]\nonumber \\
    &\times \Delta_{12}(\bm r_1, \sigma_1, \bm r_2, \sigma_2,\bm r_1', \sigma_1',\bm r_2, \sigma_2; t).
    \label{eq:C1}
\end{align}
This collision operator is non-local and describes the coupling and transfer of the electrons between $(\bm r_1, \sigma_1)$ and $(\bm r_1', \sigma_1')$ due to correlations with surrounding electrons.\\
The equations of motion of the natural orbitals $| \eta_j(t)\rangle$ follow from the off-diagonal ($i\neq j$) elements of Eq.~\ref{eq:1rdm_eom_nat} and Eq.~\ref{eq:1rdm_dots} as
\begin{align}\label{eq:eom_nat_orbs}
i\partial_t |\eta_{j}(t)\rangle &= \hat h_1|\eta_{j}(t)\rangle + \hat V_1 (t)|\eta_{j}(t)\rangle \nonumber \\
& + \sum_{i\neq j} \left( \hat{W}^i_i(t)\eta_i|\eta_j\rangle - \hat{W}^i_j(t) \eta_i |\eta_i\rangle \right) \nonumber \\
& + \sum_{i\neq j} |\eta_{i}(t)\rangle \frac{\langle\eta_{i}(t)|\text{Tr}_2[\hat W_{12}(t),\Delta_{12}(t)]|\eta_{j}(t)\rangle}{\eta_{j}(t)-\eta_{i}(t)}.
\end{align}
In the present case of strong-field driving of atoms, the first line on the right-hand side of Eq.~\ref{eq:eom_nat_orbs} includes the kinetic energy, the nuclear Coulomb potential, and the interaction with the laser field. The second line represents the direct and exchange mean-field contributions in which the conventional two-electron Coulomb matrix elements 
\begin{align}
    W_{jl}^{ik}(t) &=  \Bigl \langle \eta_i(t) \eta_k(t) \Big| \frac{1}{|\bm r_1 - \bm r_2|} \Big| \eta_j(t) \eta_l(t) \Bigr  \rangle \nonumber \\
    &= \delta^{\sigma_i}_{\sigma_j}\delta^{\sigma_k}_{\sigma_l} \int d^3r_1d^3r_2\, \nu_{q_i}^*(\bm{r}_1;t)\nu_{q_k}^*(\bm{r}_2;t)\nonumber \\
    &\times \frac{1}{|\bm{r}_1-\bm{r}_2|} \nu_{q_j}(\bm{r}_1;t)\nu_{q_l}(\bm{r}_2;t),
\end{align}
appear as effective one-electron operators whose spatial and spin representation reads
\begin{equation}
    W_j^i(\bm{r}_1;t) = \delta^{\sigma_i}_{\sigma_j} \int d^3r_2 \nu_{q_i}^*(\bm{r}_2;t) \frac{1}{|\bm{r}_1-\bm{r}_2|} \nu_{q_j}^*(\bm{r}_2;t).
\end{equation}
The third line of Eq.~\ref{eq:eom_nat_orbs} contains the contributions from the collision operator (Eq.~\ref{eq:eom_nat_occ}) unambiguously signifying correlation effects. (We note that terms in Eq.~\ref{eq:eom_nat_orbs} where the denominator vanishes do not effectively contribute.) Comparison between Eqs.~\ref{eq:eom_nat_occ} and \ref{eq:eom_nat_orbs} shows that temporal variations of the occupation numbers are exclusively due to the collision operator, i.e.~due to correlation effects ($\propto\Delta_{12}$) and obviously absent in any mean-field description, while the time evolution of the natural orbitals is governed by a combination of both one-particle and two-particle interactions. The time dependence of the occupation numbers is therefore a particularly sensitive probe of dynamical electron-electron correlations during strong-field driving. We note that $\hat{C}_1$ bears some resemblance to the so-called time-dependent transition density matrix (TDM) employed in TDDFT, mostly in the perturbative regime \cite{furche_density_2001, li_time-dependent_2011}. However, while the TDM depends on all non-local couplings encoded in the 1RDM including those due to mean-field effects, $\hat{C}_1$ selects exclusively those couplings and transitions that are caused by two-particle correlations.
\section{Measures of correlations}\label{sec:corr_meas}
Within the context of quantum information theory, the 1RDM has been extensively used to quantify entanglement in many-body quantum systems \cite{nielsen_quantum_2010}. Frequently employed measures include the purity $P=\text{Tr}D_1^2$, often also referred to as the linear entropy \cite{bose_mixedness_2000}, and the von Neumann entropy (or entanglement entropy) \cite{nielsen_quantum_2010} 
\begin{equation}\label{eq:S1}
    S_1(t) = -\text{Tr}\left[D_1(t)\ln D_1(t)\right] .
\end{equation}
From the spectral representation of $D_1$ (Eq.~\ref{eq:1rdm_eig}), it is obvious that both the linear entropy and the von Neumann entropy are completely characterized by the eigenvalues of $D_1$
\begin{equation}\label{eq:purity_eig}
    P(t) = \sum_{j}\eta_j(t)^2,
\end{equation}
\begin{equation}\label{eq:entropy_eig}
    S_1(t) = -\sum_{j}\eta_j(t)\ln{\eta_j(t)}.
\end{equation}
In turn, as the temporal evolution of occupation numbers is exclusively determined by correlation effects beyond mean-field (see Eq.~\ref{eq:eom_nat_occ}), these measures are equally well suited to quantify the temporal variation of dynamical correlations. In fact, entanglement and correlations have become the focus of recent studies of photoionization \cite{bourassin-bouchet_quantifying_2020,vrakking_control_2021,koll_experimental_2022,nandi_generation_2024, jiang_time_2024, ishikawa_control_2023, shen_coherent_2025} and portions of $D_1$ have been directly measured for photoelectrons in a RABBIT-like setting \cite{laurell_continuous-variable_2022, laurell_measuring_2023}.\\
Focusing in the following on the von Neumann entropy (Eq.~\ref{eq:S1}), $S_1$ is well suited to characterize the statistical distribution of occupation numbers. For an uncorrelated state of indistinguishable fermions, i.e.~a Slater determinant, the eigenvalues of the 1RDM are restricted to $\eta_j=1$ or $0$ and $S_1$ vanishes ($S_1=0$). In particular, mean-field approximations such as TDHF or TDDFT will give $S_1=0$ for all times. For correlated states $S_1>0$. Accordingly, dynamical (i.e.~time-dependent) correlations due to strong driving by the laser field can be quantified by the time-dependent variation of the von Neumann entropy $\delta S_1(t) = S_1(t)-S_1(0)$ where $S_1(0)$ quantifies the initial correlations of the many-electron (ground) state prior to the onset of the laser field.\\
Alternatively to $S_1$, the two-particle cumulant $\Delta_{12}$ itself (see Eqs.~\ref{eq:cum_expansion} and \ref{eq:eom_nat_occ}) can be employed as a measure of correlations as it explicitly represents two-particle correlations. We therefore introduce as an alternative measure the magnitude of $\Delta_{12}$ determined by the Hilbert-Schmidt norm (or Schatten $2$-norm) $||\Delta_{12}||$. In our applications to multi-electron atoms we will compare the temporal variation of $S_1$ with that of $||\Delta_{12}||$. We find that these two measures yield qualitatively remarkably similar results. We note that the sub-cycle variations of the energies of the natural orbitals and of the spatial energy densities have been previously suggested as a tool to distinguish dynamical correlations from mean-field effects in molecules \cite{kato_time-dependent_2009}.
\section{Correlations and HHG in multi-electron atoms}\label{sec:hhg_corr}
\subsection{Simulation methods}\label{subsec:sim_methods}
\begin{figure}[t]
    \centering
    \includegraphics[width=\columnwidth]{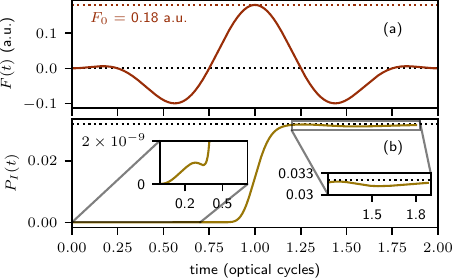}
    \caption{(a) Two-cycle pulse $F(t)$ oriented along the $z$ axis with peak field strength $F_0=0.18$ for helium. (b) The single-ionization probability $P_I(t)$ (Eq.~\ref{eq:single_ion}) for helium as a function of time for $R_I = 20$ on a linear scale, small deviations from the monotonic increase are shown on an expanded linear scale in the sub-frames.}
    \label{fig:ionization}
\end{figure}

In this section, we present first applications of the quantitative description of correlation dynamics to strong-field driving and HHG in multi-electron systems. As prototypical examples we consider the noble gas atoms helium and neon and the alkaline earth atoms beryllium and magnesium (see Tab.~\ref{tab:parameters}).
\begin{table}
    \centering
    \begin{tabular}{c|c c | c c}
        \hline
        \hline
         & He & Ne & Be & Mg\\
        \hline
         $I_p$ in $\text{eV}$ & $24.6$ & $21.6$ & $9.32$ & $7.65$\\
         $\lambda$ in $\text{nm}$ & $800$ & $800$ & $2000$ & $3200$\\
         $I_0$ in $\text{W}/\text{cm}^2$ & $1.15\times 10^{15}$ & $1\times 10^{15}$ & $5\times 10^{13}$ & $1.6\times 10^{13}$\\
         $\gamma$ & $0.425$ & $0.426$ & $0.501$ & $0.502$\\
         $E_\text{cut-off}$ in $\text{eV}$ & $242$ & $211$ & $68.5$ & $56.1$\\
         $P_I$ & $0.03$ & $0.07$ & $0.10$ & $0.03$\\
        \hline
        \hline
    \end{tabular}
    \caption{Parameters used in the simulation of strong-field driving of He, Ne, Be, and Mg by a two-cycle laser pulse chosen to yield comparable Keldysh parameters. $I_p$: ionization potential \cite{NIST_ASD}, $\lambda$: wavelength, $I_0$: intensity of the driving field,  $\gamma$: Keldysh parameter (Eq.~\ref{eq:Keldysh}), $E_\text{cut-off}$: HHG cut-off energy calculated with the classical formula Eq.~\ref{eq:E_cut}, and $P_I$: ionization probability at the conclusion of the pulse.}
    \label{tab:parameters}
\end{table}
As the driving field we use a two-cycle laser pulse given by 
\begin{align}
    A_z(t) &= - c\int_{-\infty} ^t F(t') dt',
\end{align}
\begin{align}\label{eq:el-field}
    F(t) = 
    \begin{cases}
        0  & t\leq 0 \\
        F_0\cos{(\omega t)}\sin^2{(\frac{\omega t}{4})} & 0<t\leq 2T\\
        0 & t> 2T,
    \end{cases}
\end{align}
linearly polarized in the z-direction with intensities $I=cF_{0}^2/8\pi$ (Tab.~\ref{tab:parameters}). The choice of this two-cycle laser field [see Fig.~\ref{fig:ionization} (a)] is motivated by the fact that for this field we can identify and analyze two different ionization scenarios triggered within the same pulse.
One sequence starts with the ionization by a still (relatively) weak field near the first extremum [a minimum in Fig.~\ref{fig:ionization} (a)] followed by acceleration of the liberated electron by a stronger field near the second extremum [a maximum in Fig.~\ref{fig:ionization} (a)]. The other sequence starts with the ionization burst triggered by the strong field near this maximum followed by acceleration by the already weaker field near the third extremum. Contributions from these two processes can be well distinguished from each other in the resulting HHG spectrum as their respective yields are, as suggested by the three-step model, governed by the product of the ionization rate $\dot N_\text{ion}$ at the ionization time $t_\text{ion}$ and the recombination probability $P_\text{rec}$ at the time $t_\text{rec}$ 
\begin{equation}\label{eq:yield_estimate}
    I^\text{HHG} \propto \dot{N}_\text{ion}(t_\text{ion})P_\text{rec}(t_\text{rec}).
\end{equation}
The first sequence can reach the cut-off energy
\begin{equation}\label{eq:E_cut}
    E_\text{cut-off} = 3.17 U_p + I_p,
\end{equation}
corresponding to the ponderomotive energy $U_p = \frac{F_0^2}{4\omega^2}$ of the peak field and $I_p$ the ionization energy. However, it has a lower yield than the second sequence since the tunnel ionization rate $\dot{N}_I$ is still relatively small near the first field extremum. In turn, the second sequence can only reach a much lower energy since the electron is accelerated by an already diminished field. The present choice of the field parameters ($F_0$ and $\omega$) is furthermore optimized such that significant one-electron ionization takes place while two-electron ionization remains negligible. With this choice, specific correlation effects that can be attributed to double ionization (see e.g.~\cite{li_high-order_2019}) are largely suppressed.\\
In order to realize comparable scenarios for the different atomic species, we have chosen the laser parameters such that the corresponding Keldysh parameter 
\begin{equation}\label{eq:Keldysh}
    \gamma = \sqrt{\frac{I_p}{2U_p}}
\end{equation}
is similar ($\gamma\approx 0.5$) for all four cases despite the large differences in the ionization potentials $I_p$ (Tab.~\ref{tab:parameters}).\\
The one-electron ionization probability is calculated from 
\begin{equation}\label{eq:single_ion}
    P_I(t) = \int_{|\bm{r}|> R_I} d^3r\,\rho(\bm r,t),
\end{equation}
with the electron density given by the diagonal elements of the 1RDM in spatial representation $\rho(\bm r,t) = \sum_{\sigma} D_1(\bm r, \sigma,\bm r, \sigma;t)$. (For numerical values of $R_I$ see Appendix~\ref{app:numerics}.) 
Since the double ionization probability remains negligible in the present simulations, the ionization rate in Eq.~\ref{eq:yield_estimate} can be obtained from 
\begin{equation}\label{eq:ion_rate}
    \dot N_\text{ion}(t) = \frac{d}{dt}P_I(t).
\end{equation}
While Eq.~\ref{eq:single_ion} can serve as a reliable measure for the ionization probability only for sufficiently large $R_I$ and after the conclusion of the pulse, the time-dependent ionization rate $\dot N_\text{ion}(t)$ can be used to approximately characterize the ionization burst during tunneling or over-the-barrier ionization also in the presence of the laser field.\\

For the present numerical results we determine the ``exact" (i.e.~converged) time-dependent $N$-electron wavefunction $|\Psi(t)\rangle$ from the solution of the Schr\"odinger equation (Eq.~\ref{eq:tdse}) using the MCTDHF method \cite{caillat_correlated_2005, kato_time-dependent_2004, nest_multiconfiguration_2005, haxton_multiconfiguration_2011, karabanov_accuracy_2011, sato_time-dependent_2016}. Briefly, the ansatz for the wavefunction is given by
\begin{equation}\label{eq:psi_ci_ansatz}
    |\Psi(t)\rangle = \sum_J C_J(t)|\Phi_J(t)\rangle,
\end{equation}
with the time-dependent configuration-interaction (CI) coefficients $C_J(t)$, and $|\Phi_J(t)\rangle$ the corresponding Slater-determinant basis built from $2M$ time-dependent spin orbitals $\{|\xi_i(t)\rangle\}$ 
with $|\xi_i\rangle = |\varphi_{q_i}\rangle \otimes |\sigma_i\rangle$ and $i = (q_i, \sigma_i) = (nl_{m_l}, \uparrow)$ for $i<M$ and $(nl_{m_l}, \downarrow)$ for $i>M$ (where we use the standard spectroscopy labeling for the orbitals according to their zero-field designation). The sum in Eq.~\ref{eq:psi_ci_ansatz} runs over all possible Slater determinants built from the $2M$ spin orbitals. The fact that within the MCTDHF method the orbitals can self-consistently adapt to external driving is crucial for simulations of the multi-electron dynamics in strong fields. Any stationary basis would require a prohibitively large number of configurations in Eq.~\ref{eq:psi_ci_ansatz}. For the orbitals entering the Slater-determinant basis a partial-wave expansion
\begin{equation}\label{eq:working_orbitals}
    \varphi_{k,m}(r,\theta, \phi, t) = \sum_{l=0}^{l_\text{max}}R^l_{k,m}(r,t)Y_{lm}(\theta, \phi),
\end{equation}
with $l_\text{max}$ depending on the atom varying from $47$ to $95$ (for details see App.~\ref{app:numerics}) is employed accounting for the $l$-mixing by the driving field. For linear polarization along the $z$-axis (see Eq.~\ref{eq:el-field}) the magnetic quantum number $m$ is preserved.\\
From the propagated wavefunction $|\Psi(t)\rangle$ the reduced density matrix (Eq.~\ref{eq:1rdm}), the corresponding measures of correlations (Eqs.~\ref{eq:S1} and Eq.~\ref{eq:cum_expansion}), and the HHG spectra (Eq.~\ref{eq:Larmor}) are deduced.
\subsection{Dynamics of correlations}\label{subsec:corr_dyn}
\begin{figure}
    \centering
    \includegraphics[width=\columnwidth]{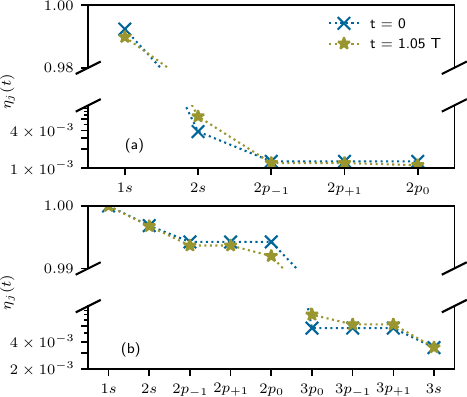}
    \caption{Snapshots of occupation numbers for natural orbitals with spin-up of noble gas atoms He (a) and Ne (b) prior to the pulse ($t=0$) and near the peak of the pulse strength ($t=1.05\ \text{T}$) (field parameters as described in the text). The natural orbitals are labeled according to their zero-field designation, spin-up and spin-down occupation numbers are identical. Note the change in scale to display small occupation numbers above the Fermi edge.}
    \label{fig:fermi_He_Ne}
\end{figure}

\begin{figure}
    \centering
    \includegraphics[width=\columnwidth]{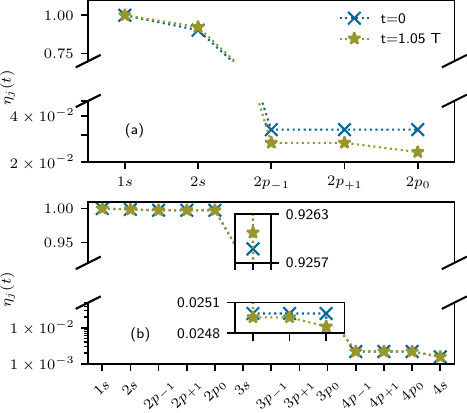}
    \caption{Same as Fig.~\ref{fig:fermi_He_Ne} but for the alkaline earth atoms Be (a) and Mg (b) where the sub-frames are zoom-ins for the HONO and LUNO orbitals to enhance the visibility of small changes.}
    \label{fig:fermi_Be_Mg}
\end{figure}

\begin{figure}
    \centering
    \includegraphics[width=\columnwidth]{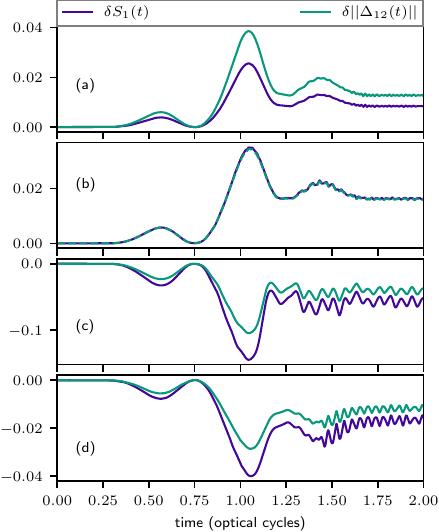}
    \caption{Measures for dynamical correlations. Variation of the von Neumann entropy $\delta S_1(t) = S_1(t)-S_1(0)$ and of the norm of the two-particle cummulant $\delta ||\Delta_{12}(t)|| = ||\Delta_{12}(t)||-||\Delta_{12}(0)||$ for (a) He, (b) Ne, (c) Be and (d) Mg. Note the sign change between (a,b) and (c,d).}
    \label{fig:measures}
\end{figure}

The time-dependent variation of the ionization probability $P_I(t)$ (Eq.~\ref{eq:single_ion}) driven by the pulse displayed in Fig.~\ref{fig:ionization} (a), is shown for helium in Fig.~\ref{fig:ionization} (b). $P_I(t)$ is an almost monotonically increasing function during the central half cycle with the half-way point at the center of the pulse, $P_I(1.0\ \text{T})=P_I(\infty)/2$ with $P_I=P_I(\infty)=0.03$. Small modulations due to the first ionization burst near $t=0.3\ \text{T}$ and during rescattering or recombination near $t=1.5\ \text{T}$ become visible only on an expanded scale [Fig.~\ref{fig:ionization} (b)]. Structurally identical results (not shown) are found for neon with $P_I=0.07$, beryllium with $P_I=0.10$, and magnesium with $P_I=0.03$.\\
Changes in the occupation numbers of the natural orbitals under the influence of strong-field driving (Eq.~\ref{eq:eom_nat_occ}) are displayed for helium in Fig.~\ref{fig:fermi_He_Ne} (a) where we compare the initial (ground) state occupation numbers at $t=0$ with those close to the peak field [i.e.~at $t=1.05\ \text{T}$ corresponding to the maximum in the correlation measures shown in Fig.~\ref{fig:measures} (a) below]. Note that for convenience the labeling of the natural orbital occupation numbers in Fig.~\ref{fig:fermi_He_Ne} and the following figures refers to their initial ground-state designation. In the presence of the driving field, the radial shapes of the natural orbitals as well as their $l$ characteristics are strongly perturbed. Since the initial state includes ground-state correlations, the ``Fermi edge" in the occupation numbers is not a perfect step function at $t=0$ (compare with Eq.~\ref{eq:1rdm_mean_field}) but (slightly) smoothed. Under the influence of the strong-field driving the Fermi edge gets further smoothed as the population is shifted to additional higher orbitals. While the latter appears to be, at first glance, intuitively obvious, it represents, in fact, a non-trivial dynamical correlation effect as changes in the occupation of the natural orbitals are exclusively driven by the two-particle correlator (Eq.~\ref{eq:eom_nat_occ}) while strong-field and mean-field effects are, to leading order, already included in the (occupied) natural orbitals themselves (Eq.~\ref{eq:eom_nat_orbs}). A qualitatively very similar pattern emerges for neon [Fig.~\ref{fig:fermi_He_Ne} (b)]. By contrast, for the alkaline earth atoms beryllium [Fig.~\ref{fig:fermi_Be_Mg} (a)] and magnesium [Fig.~\ref{fig:fermi_Be_Mg} (b)] a markedly and surprisingly different time evolution of the natural orbital occupation is observed. The Fermi edge of the occupation numbers gets sharper in the presence of the strong field rather than smoother.\\
These trends observed by the snapshots at $t=1.05\ \text{T}$ are confirmed by the time-dependence of the correlation measures, the field-induced temporal variation of entropy $S_1$, $\delta S_1(t) = S_1(t)-S_1(0)$ and of the variation of the norm of the cumulant $\delta ||\Delta_{12}(t)||=||\Delta_{12}(t)||-||\Delta_{12}(0)||$ (Fig.~\ref{fig:measures}). While the von Neumann entropy and the norm of the two-particle correlator both increase - even though non-monotonically - during the strong-field driving of helium and neon, they decrease for beryllium and magnesium. For all systems investigated, the time evolution of these two complementary measures of correlations closely agree with each other supporting the notion of the close correspondence between entanglement and correlations for strongly driven fermionic systems. The local extrema of the correlation measures are well synchronized with the extrema of the field amplitude. For the noble gases the dynamical correlation has local maxima near the field extrema while for alkaline earth atoms these correspond to local minima.\\

\begin{figure}
    \centering
    \includegraphics[width=\columnwidth]{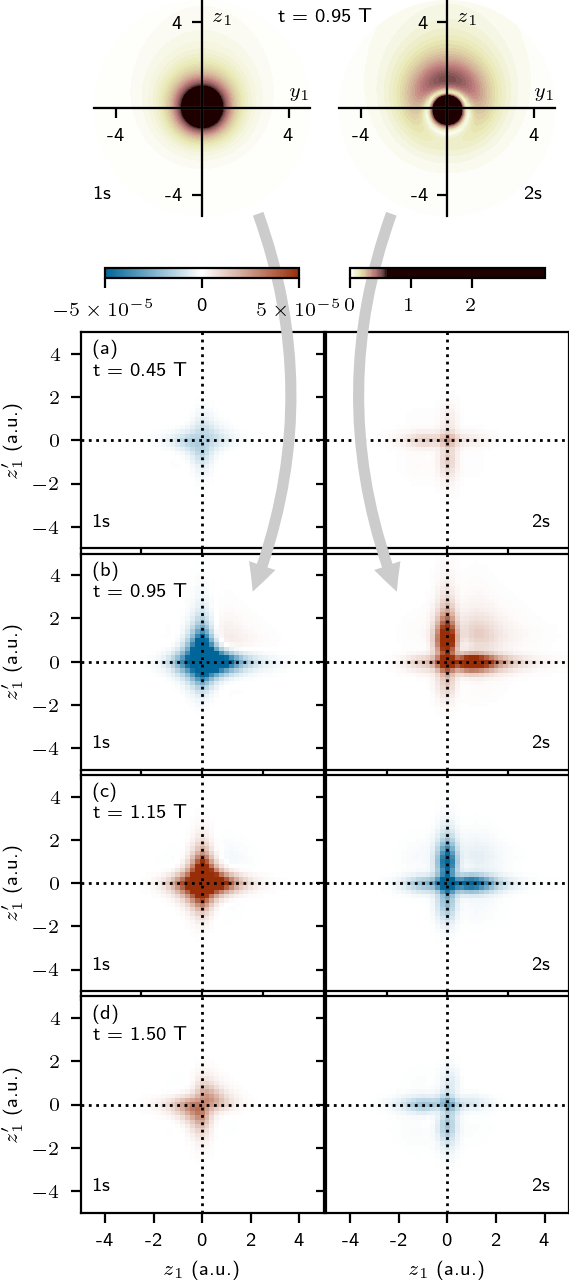}
    \caption{Spatial distribution of $\text{Im}(\eta_j|\hat{C}_1(z_1,z_1')|\eta_j)$ in the $z_1 - z_1'$ plane for the HONO (left) and the LUNO (right) of He at (a) $t = 0.47~\text{T}$, (b) $t = 0.95~\text{T}$, (c) $t = 1.15~\text{T}$, and (d) $t = 1.50~\text{T}$. Source regions with positive values of $\text{Im}(\eta_j|\hat{C}_1(z_1,z_1')|\eta_j)$ in red, sinks in blue. The two top frames display the magnitude of the spatial HONO and LUNO orbitals in the $x_1=0$ plane $|\nu_{q_j}(x_1=0,y_1,z_1;t)|$ at time $t=0.95~\text{T}$.}
    \label{fig:C1_He}
\end{figure}

A simple qualitative explanation of the different correlation dynamics for noble-gas and alkaline-earth atoms refers to the electronic structure of the initial ground state: for noble-gas atoms the strong-field excitation and ionization opens their initial closed-shell electronic structure thereby naturally increasing correlation effects. By contrast for alkaline-earth atoms with two relatively weakly bound electrons outside the noble-gas closed shell configuration featuring (relatively) strong correlations in their initial ground state prior to the interaction with the laser field, strong-field ionization of one of these electrons results in an increased phase-space (i.e.~position and momentum) separation between this electron pair thereby reducing electron correlations.\\
A more detailed quantitative analysis requires the investigation of the properties of the collision operator (Eq.~\ref{eq:eom_nat_occ}) governing the influence of two-particle correlations on the natural orbitals and their occupation (Eq.~\ref{eq:eom_D1}). Taking into account that during the time evolution in the laser field both $\hat{C}_1(t)$ and the natural orbitals $|\eta_j(t)\rangle$ acquire complex phases, the time differential change in occupation number (Eq.~\ref{eq:eom_nat_occ}) is governed by the matrix element of the collision operator
\begin{align}
    &\text{Im}\langle \eta_j|\hat{C}_1|\eta_j\rangle \nonumber=
    \sum_{\sigma_1}\int d^3r_1d^3r_1'C_1(\bm r_1,\sigma_1, \bm r_1', \sigma_1)\\
    & \times \left[
    \nu_{q_j}^R(\bm r_1)\nu_{q_j}^R(\bm r'_1) + 
    \nu_{q_j}^I(\bm r_1)\nu_{q_j}^I(\bm r'_1)\right]
    \label{eq:F1}
\end{align}
with $\nu_{q_j}^R(\bm r_1)=\text{Re}\,\nu_{q_j}(\bm r_1)$ and $\nu_{q_j}^I(\bm r_1)=\text{Im}\,\nu_{q_j}(\bm r_1)$. To explore the local behavior of the collision operator along the field direction ($z$) we rewrite this matrix elements as 
\begin{align}\label{eq:Im_C_1}
    \text{Im}\langle \eta_j|\hat{C}_1|\eta_j\rangle
    =\int dz_1\int dz_1'\ \text{Im}(\eta_j|\hat{C}_1(z_1,z_1') |\eta_j)
\end{align}
where the rounded brackets stand for the integration over the transverse coordinates $(x_1,y_1,x_1',y_1')$ only. We study the spatially resolved time variation of $\text{Im}(\eta_j|\hat{C}_1(z_1,z_1')|\eta_j)$. The physical interpretation of the off-diagonal elements of the collision operator $\text{Im}(\eta_j|\hat{C}_1(z_1,z_1')|\eta_j)$ is that of a one-particle transition rate per unit time and unit area for transferring population of the natural orbital $j$ when one (projected) coordinate of the natural orbital along the direction of the driving field is at $z_1$ and the other at $z_1'$. We focus on the matrix elements for the two orbitals closest to the Fermi edge set at $\eta_j=0.5$, with the highest (i.e.~strongly) occupied natural orbital (HONO), $\eta_j>0.5$, and the lowest unoccupied (more precisely, weakly occupied) natural orbital (LUNO), $\eta_j<0.5$ and take snapshots of their distributions in the $z_1-z_1'$ plane at the instants of time when the population transfer between them has local maxima (e.g.~$0.45\ \text{T}$, $0.95\ \text{T}$, $1.15\ \text{T}$ and $1.50\ \text{T}$ for He).\\
Fig.~\ref{fig:C1_He} shows $\text{Im}(\eta_j|\hat{C}_1(z_1,z_1')|\eta_j)$ for helium for which the first two natural orbitals are the HONO ($q_1=1$s) and the LUNO ($q_2=2$s). 
\begin{figure}
    \centering
    \includegraphics[width=0.99\columnwidth]{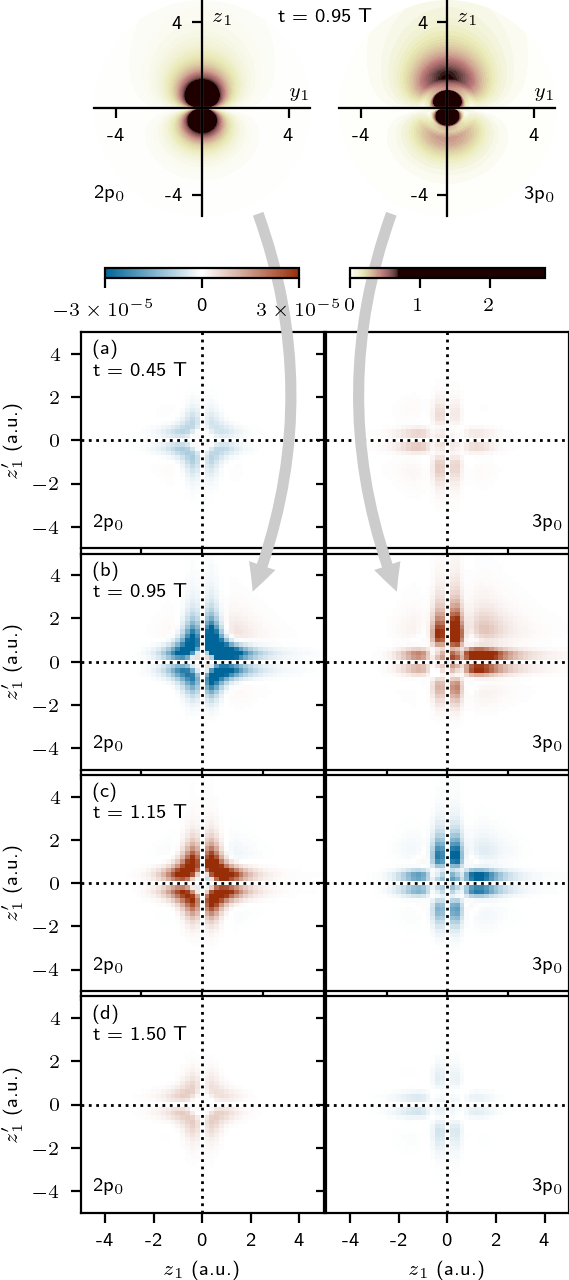}
    \caption{As Fig.~\ref{fig:C1_He} but for Ne.}
    \label{fig:C1_Ne}
\end{figure}
\begin{figure}
    \centering
    \includegraphics[width=0.99\columnwidth]{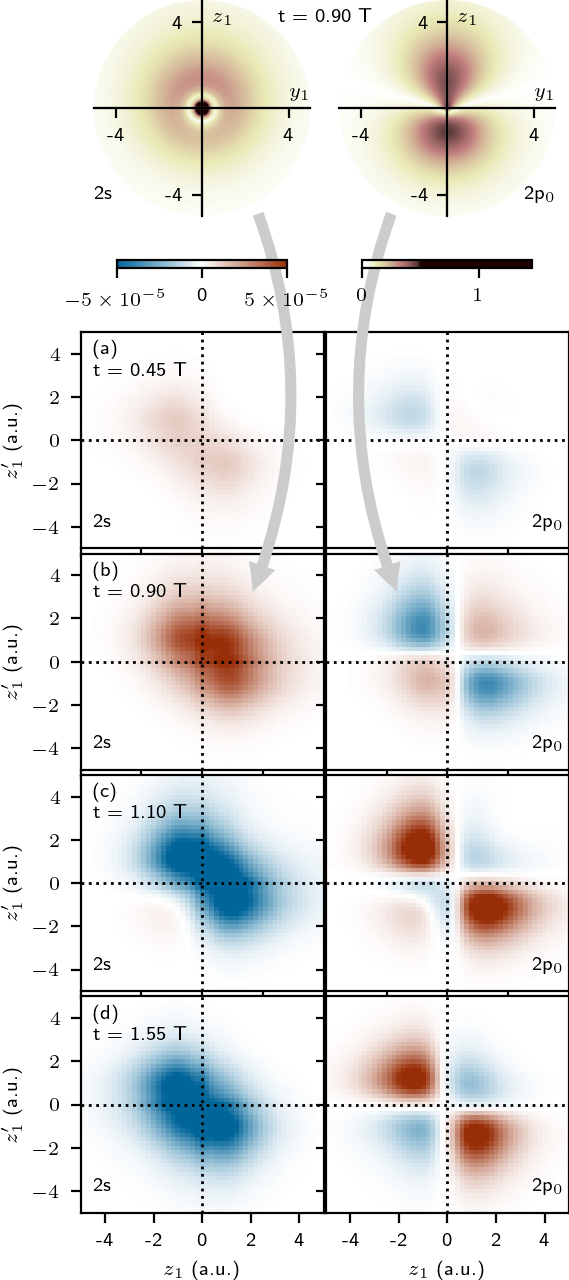}
    \caption{As Fig.~\ref{fig:C1_He} but for Be and at (a) $t = 0.45~\text{T}$, (b) $t = 0.90~\text{T}$ , (c) $t = 1.10~\text{T}$ , and (d) $t = 1.55~\text{T}$ .}
    \label{fig:C1_Be}
\end{figure}
\begin{figure}
    \centering
    \includegraphics[width=0.99\columnwidth]{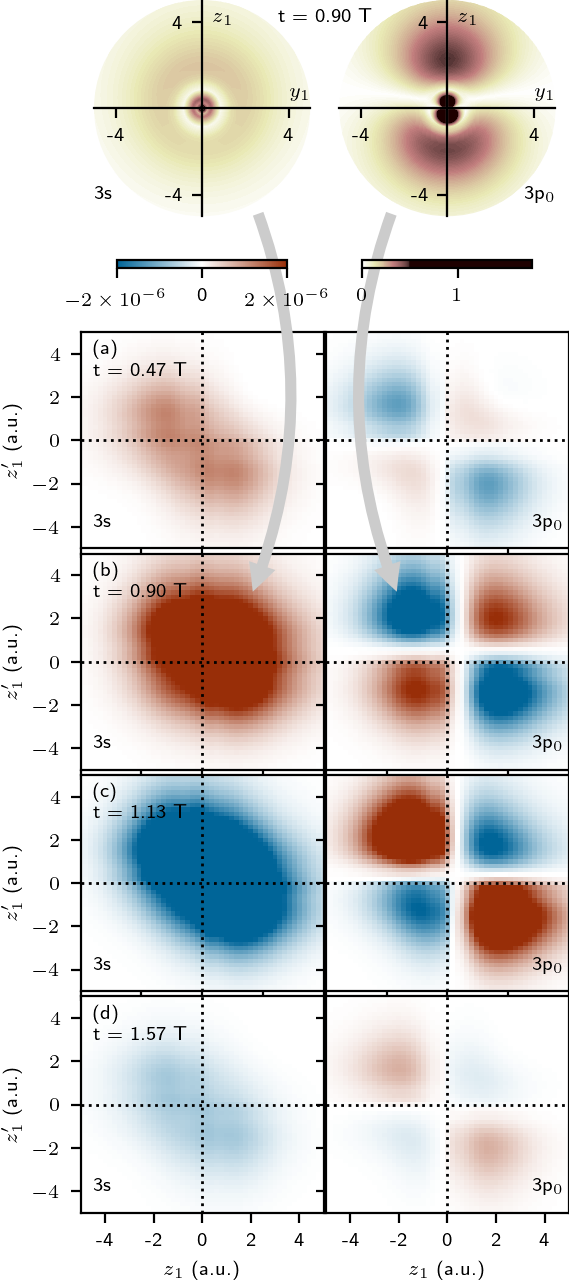}
    \caption{As Fig.~\ref{fig:C1_He} but for Mg and at (a) $t = 0.47~\text{T}$, (b) $t = 0.90~\text{T}$, (c) $t = 1.13~\text{T}$, and (d) $t = 1.57~\text{T}$.}
    \label{fig:C1_Mg}
\end{figure}
For simplicity we label them by their ground-state designation in the absence of the laser field even though their angular and radial nodal structures are distorted by the driving field. On the rising flank of the pulse envelope (at $t=0.45\ \text{T}$ and $t=0.95\ \text{T}$) the population flux is directed from the HONO ($1$s) to the LUNO ($2$s) as expected from the smoothing of the Fermi edge [Fig.~\ref{fig:fermi_He_Ne} (a)]. This correlation-induced flux originates predominantly from the region in the $z_1-z_1'$ plane where both coordinates are on the same side of the atom near the tunneling barrier, i.e.~in the third quadrant at $t=0.45\ \text{T}$ and in the first quadrant at $t=0.95\ \text{T}$ (Fig.~\ref{fig:C1_He}). Tunneling (or over-barrier) excitation and ionization dynamics thus enhances the correlation induced transfer at closely spaced coordinate pairs in helium. On the falling flank of the envelope (at $t=1.15\ \text{T}$ and $t=1.50\ \text{T}$) the population flux originates again preferentially from regions on the same side of the atom near the barrier (first quadrant at $t=1.15\ \text{T}$ , fourth quadrant at $t=1.50\ \text{T}$), but the direction of the flux is now reversed, i.e.~ from the LUNO to the HONO. The latter signifies the field-driven return to the ionic core of the ionized wavepacket where a small portion of it will recombine. A qualitatively similar pattern emerges for neon (Fig.~\ref{fig:C1_Ne}). The $z_1-z_1'$ distribution of $\text{Im}(\eta_j|\hat{C}_1|\eta_j)$ displays in this case a more complex nodal structure with nodal planes $z_1=0$ and $z_1'=0$ of the $2$p$_0$ and $3$p$_0$ orbitals visible. In addition, also remnants of the radial node of the $3$p$_0$ orbital are recognizable, similar to those of the $2$s orbitals for helium (Fig.~\ref{fig:C1_He}). Overall, the temporal evolution of the matrix element of the collision operator for the HONO and LUNO closely match that of helium, underscoring the structurally similar response of noble gas atoms to strong driving. Both the direction of flux between HONO and LUNO as well as the preferred region of its origin in the $z_1-z_1'$ plane, on the same side and near the lowered tunneling barrier, agree.\\
Drastically different $z_1-z_1'$ distributions of the matrix elements of the collision operator are found for the alkaline-earth atoms beryllium (Fig.~\ref{fig:C1_Be}) and magnesium (Fig.~\ref{fig:C1_Mg}). The direction of flux between HONO and LUNO in the alkaline-earth atoms is opposed to that in the noble-gas atoms. As strong-field ionization increases on the rising flank of the pulse, the flux is directed from the LUNO to the HONO while on the falling flank the flux is heading back towards the LUNO. Even more surprisingly and importantly, the spatial distributions are entirely different: sinks and sources of the flux are predominantly localized on opposite sides of the atom (i.e.~in the second and fourth quadrant) with only one coordinate near the tunneling barrier and larger distances between $z_1$ and $z_1'$. This is a consequence of the fact that the initial state of the two electrons outside the closed shell is both more strongly correlated and more delocalized. Furthermore, the position of the peaks of the sinks and sources are only slightly affected by the strong field displaying only slight shifts along the direction of the field ionization, i.e.~along the diagonal coordinate $z_1+z_1'$. \\
Unlike for noble gases, the LUNO for alkaline-earth atoms features in addition to the dominant flux contribution from the second and fourth quadrant also a weak but clearly visible contribution of flux in the opposite direction stemming from the first and third quadrant (Figs.~\ref{fig:C1_Be} and \ref{fig:C1_Mg}). This pattern originates most likely from the change of parity under reflection on $R_z$ ($z \rightarrow -z$) when transitioning between the $R_z$-even HONO ($2$s in Be, $3$s in Mg) and the $R_z$-odd LUNO ($2$p$_0$ in Be, $3$p$_0$ in Mg). The influence of the nodal plane ($z_1=0$) remains also clearly visible in the $x_1=0$ cut of the magnitude of the spatial orbitals $|\nu_{q_j}(x_1=0,y_1,z_1;t)|$ of the HONO and LUNO displayed in the top frames of Figs.~\ref{fig:C1_He}-\ref{fig:C1_Mg}. For noble gases, the (absence of the) nodal plane is preserved in the transition between HONO and LUNO while for alkaline-earth atoms, the nodal plane (dis)appears when switching between the HONO and the LUNO.
\FloatBarrier
\subsection{HHG spectra}\label{subsec:hhg_spec}
\begin{figure*}
    \centering
    \includegraphics[width=\textwidth]{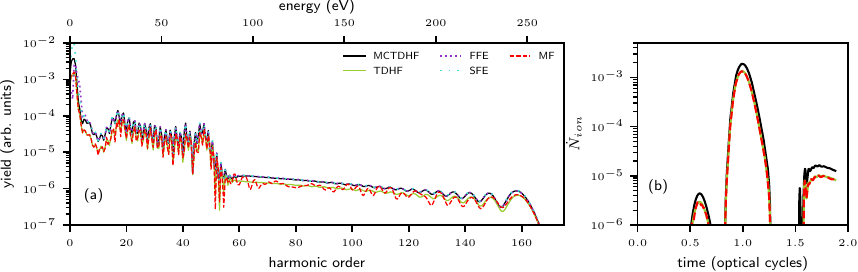}
    \caption{He driven by a strong two-cycle laser field with parameters according to Tab.~\ref{tab:parameters}. (a) HHG spectrum obtained by MCTDHF (black), TDHF (green), MF (red), FFE (purple), and SFE (blue). Note that FFE and SFE nearly coincide with MCTDHF. (b) Ionization rate $\dot{N}_\text{ion}$ as a function of time (Eq.~\ref{eq:ion_rate}), same color code as in (a).}
    \label{fig:HHG_He}
\end{figure*}

\begin{figure*}
    \centering
    \includegraphics[width=\textwidth]{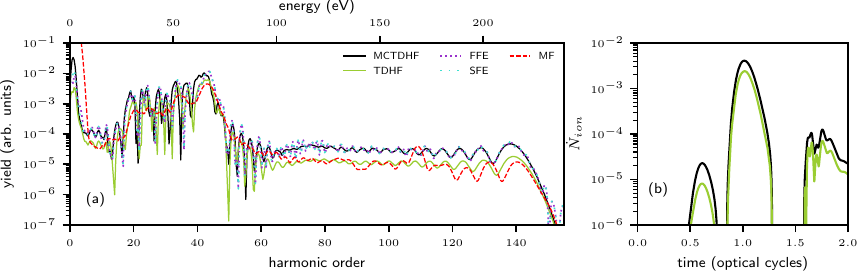}
    \caption{Ne driven by a strong two-cycle laser field with parameters according to Tab.~\ref{tab:parameters}. (a) HHG spectrum obtained by MCTDHF (black), TDHF (green), MF (red), FFE (purple), and SFE (blue). Note that FFE and SFE nearly coincide with MCTDHF. (b) Ionization rate $\dot{N}_\text{ion}$ (Eq.~\ref{eq:ion_rate}), same color code as in (a).}
    \label{fig:HHG_Ne}
\end{figure*}

\begin{figure*}
    \centering
    \includegraphics[width=\textwidth]{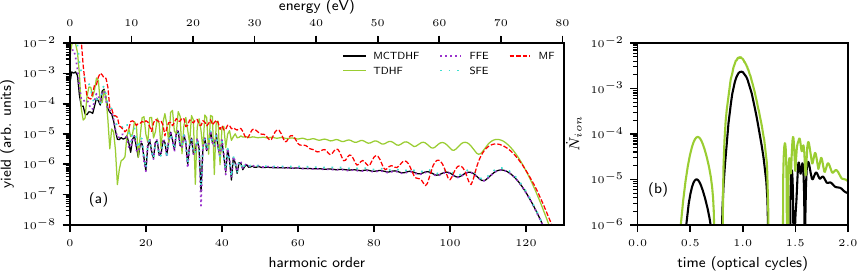}
    \caption{Be driven by a strong two-cycle laser field with parameters according to Tab.~\ref{tab:parameters}. (a) HHG spectrum obtained by MCTDHF (black), TDHF (green), MF (red), FFE (purple), and SFE (blue). Note that FFE and SFE nearly coincide with MCTDHF. (b) Ionization rate $\dot{N}_\text{ion}$ (Eq.~\ref{eq:ion_rate}), same color code as in (a).}
    \label{fig:HHG_Be}
\end{figure*}

\begin{figure*}
    \centering
    \includegraphics[width=\textwidth]{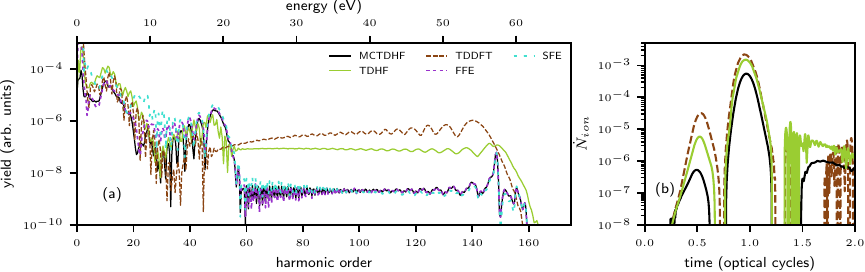}
    \caption{Mg driven by a strong two-cycle laser field with parameters according to Tab.~\ref{tab:parameters}. (a) HHG spectrum obtained by MCTDHF (black), TDHF (green), TDDFT (brown), FFE (purple), and SFE (blue). (b) Ionization rate $\dot{N}_\text{ion}$ (Eq.~\ref{eq:ion_rate}), same color code as in (a).}
    \label{fig:HHG_Mg}
\end{figure*}

We turn now to the impact of the time-dependent dynamical electron correlations on HHG. Unambiguously separating correlations effects from mean-field effects in a strong-field scenario has remained a challenge, which we attempt to address in the following. We emphasize that we focus here on the single-atom response omitting from the results propagation effects and ensemble averages which need to be included for HHG spectra of a realistic gas target.\\
The most straight-forward approach is the direct comparison between the HHG spectra from a calculation that fully includes correlations taken to be in the following from the MCTDHF method and those of well-established mean-field methods such as TDHF which excludes correlations [see Fig.~\ref{fig:HHG_He}-\ref{fig:HHG_Mg} (a)], or TDDFT [see Fig.~\ref{fig:HHG_Mg} (a)] which includes correlations only indirectly via an effective one-particle exchange-and-correlations potential. An example of such a comparison is shown in Fig.~\ref{fig:HHG_Mg} for Mg. All spectra [Fig.~\ref{fig:HHG_He}-\ref{fig:HHG_Mg} (a)] display the expected two-plateau structure. For Mg (Fig.~\ref{fig:HHG_Mg}) the higher-intensity plateau extending to about the $60^\text{th}$ harmonic originating from the strongest ionization burst [peak in $\dot{N}_\text{ion}$, see Fig.~\ref{fig:HHG_Mg} (b)] near the center of the pulse ($t\approx \text{T}$) electrons of which are accelerated by and driven back to the ion by the laser pulse with the amplitude already diminished. The low-intensity plateau extends all the way to about the $160^\text{th}$ harmonic stemming from the first ionization burst by the still relatively weak field at $t\approx 0.5\ \text{T}$ [Fig.~\ref{fig:HHG_Mg} (b)] followed by the acceleration by the stronger field around $t\approx \text{T}$. The relative height of the plateaus [Fig.~\ref{fig:HHG_Mg} (a)] is controlled by the strength of the respective ionization burst [Fig.~\ref{fig:HHG_Mg} (b)]. For example, the difference in strength between the first and second ionization burst is much larger in MCTDHF than in TDDFT. This translates into a pronounced height difference between the first and second plateau of the HHG spectrum predicted by MCTDHF as compared to TDDFT. \\

The significant discrepancies between the HHG spectra predicted by the fully correlated MCTDHF and the mean-field calculations (TDDFT or TDHF) cannot be immediately attributed to the effect of the variation in time of the correlations as already differences in the initial ground state, in particular differences in the respective ionization potentials as well as differences in the mean-field dynamics can contribute to such discrepancies. To unambiguously pinpoint the role of the dynamical change in the  correlations (increase for noble gas atoms and decrease for alkaline-earth atoms), the present formulation in terms of RDMs allows to selectively switch on or off two-particle correlations during strong-field driving. We use for this purpose the collision operator $\hat{C}_1$ in the equations of motion for the natural orbital occupation numbers (Eq.~\ref{eq:eom_nat_occ}) and the natural orbitals (Eq.~\ref{eq:eom_nat_orbs}).\\
Probing for the time-dependent dynamical correlation effects can proceed on different levels of approximation. Starting point is the same correlated initial state given by the natural orbitals $\{|\eta_j(0)\rangle\}_j$ and eigenvalues of the 1RDM, $\{\eta_j(0)\}_j$, derived from the MCTDHF calculations. Initial ground-state correlations, unlike time-dependent variations of correlations, are, thus, fully included. Orbitals and/or occupation numbers are, then, propagated without including the collision operator. Setting $\hat{C}_1=0$ (or $\Delta_{12}=0$) in both Eqs.~\ref{eq:eom_nat_occ} and \ref{eq:eom_nat_orbs}, the orbitals propagate only under the influence of the mean-field (MF) operator (first two lines on the right-hand side of Eq.~\ref{eq:eom_nat_orbs}), $|\eta_j^\text{MF}(t)\rangle$, with the initial state taken to be the exact ground-state natural orbital $|\eta_j^\text{MF}(0)\rangle =|\eta_j(0)\rangle$ and the occupation numbers remain constant. Consequently, the change of correlations over time in both the occupation numbers and the orbitals are neglected and the dipole acceleration describing the emitted radiation of the atomic ``antennas" (compare with Eq.~\ref{eq:dipole_a_nat}) is given by 
\begin{equation}\label{eq:a_MF}
    \bm{a}_\text{MF}(t) = \sum_j \eta_j(0)\langle \eta_j^\text{MF}(t)|\hat{\bm{a}}_1|\eta_j^\text{MF}(t)\rangle.
\end{equation}
It should be noted that with this choice the initial states are not eigenstates of the MF propagator. This results in additional high-frequency oscillations (``beats") unrelated to strong driving which should be removed when analyzing the spectrum. A less drastic approximation consists of neglecting only the changes in the natural orbital occupation numbers driven by the time-varying correlations, i.e.~setting the collision operator $\hat{C}_1=0$ only in Eq.~\ref{eq:eom_nat_occ}. This amounts to a frozen Fermi edge (FFE) of the initial state by neglecting its time-dependent variation while including the full evolution of the natural orbitals themselves. In this approximation the dipole acceleration is given by
\begin{equation}\label{eq:a_FE}
    \bm{a}_\text{FFE}(t) = \sum_j \eta_j(0)\langle \eta_j(t)|\hat{\bm{a}}_1|\eta_j(t)\rangle.
\end{equation}
Furthermore, in the limiting case where the shape of the initial Fermi edge is entirely neglected and the occupation numbers are set to those of the Hartree-Fock approximation, which features a sharp Fermi edge (SFE) (Eq.~\ref{eq:MF_nat_occ}) but the correlated orbital propagation is still fully included, Eq.~\ref{eq:a_FE} reduces to 
\begin{equation}\label{eq:a_SFE}
    \bm{a}_\text{SFE}(t) = \sum_{j=1}^{N_\text{occ}} \langle \eta_j(t)|\hat{\bm{a}}_1|\eta_j(t)\rangle.
\end{equation}
Comparing the HHG spectra predicted by Eqs.~\ref{eq:a_MF}, \ref{eq:a_FE}, and \ref{eq:a_SFE} with those resulting from fully correlated dynamics allows to assess in detail the influence of the time-dependent dynamical correlations. Examples are presented for helium [Fig.~\ref{fig:HHG_He} (a)], neon [Fig.~\ref{fig:HHG_Ne} (a)], beryllium [Fig.~\ref{fig:HHG_Be} (a)], and Mg [Fig.~\ref{fig:HHG_Mg} (a)]. Results for the MF model (Eq.~\ref{eq:a_MF}) are only shown for the cases where numerical convergence could be achieved [Figs.~\ref{fig:HHG_He}, \ref{fig:HHG_Ne} (a) and \ref{fig:HHG_Be} (a)]. Consistent with the relatively small correlation-induced changes of the occupation numbers relative to the sharp Fermi edge (Fig.~\ref{fig:fermi_He_Ne} and \ref{fig:fermi_Be_Mg}), the FFE and SFE models (Eq.~\ref{eq:a_FE} and \ref{eq:a_SFE}) which involve only approximations to the occupation numbers $\eta_j(t)$ while fully accounting for correlation dynamics of the natural orbitals $| \eta_j(t)\rangle$, yield HHG spectra in close agreement with the full MCTDHF calculation in most cases. A remarkable exception are the lower harmonics for the alkaline-earth atoms [Fig.~\ref{fig:HHG_Be} (a) and \ref{fig:HHG_Mg} (a)]. By contrast, when neglecting two-particle correlations in the orbital propagation by using the mean-field (MF) propagation (Eq.~\ref{eq:a_MF}) corresponding to $\hat{C}_1=0$, the resulting smoothed HHG spectra resemble that of the Hartree-Fock approximation. The drastic change in magnitude of the HHG spectrum by about one half to one order of magnitude in the second plateau region above the $50^\text{th}$ harmonic between the TDHF and the MCTDHF can thus unambiguously be assigned to the direct influence of the collision operator representing two-particle correlation effects beyond the mean field on the dynamics of the natural orbitals. Furthermore, this influence on the spectra can be traced back to a considerable extent to its influence on the sub-cycle ionization bursts $\dot{N}_\text{ion}(t)$ [see Fig.~\ref{fig:HHG_He} (b)]. Electron-electron correlations enhance the (mostly tunneling) ionization rate in the noble gases (Fig.~\ref{fig:HHG_He} for hellium and Fig.~\ref{fig:HHG_Ne} for neon) while $\dot{N}_\text{ion}(t)$ is reduced by correlations in the alkaline-earth atoms (Fig.~\ref{fig:HHG_Be} for beryllium). An analogous behavior is observed for magnesium (Fig.~\ref{fig:HHG_Mg}). This opposing trend is consistent with the opposing direction of the probability flux in the collision operator $\text{Im}(\eta_j|\hat{C}_1(z_1,z_1')|\eta_j)$ between HONO and LUNO (Figs.~\ref{fig:C1_Ne} and \ref{fig:C1_Be}). While for noble gases the flux on the rising flank of the pulse which dominantly contributes to the high-energy plateau is directed from the HONO to the LUNO, for alkaline earth atoms the direction is reversed. For magnesium we also show TDDFT results (Fig.~\ref{fig:HHG_Mg}) which resemble those of TDHF but significantly differ from fully correlated MCTDHF results.
\section{Conclusions and outlook}\label{sec:conclusions}
We have investigated the role of dynamical electron correlations in multi-electron atoms on ultrafast time scales. We have developed a one-particle reduced-density matrix (1RDM) description within which the exact $N$-electron strong-field response entering the Larmor formula can be described by an ensemble of atomic oscillators each of which given by an eigenstate and eigenvalue of the 1RDM. This formulation can be viewed as the extension of Kuchiev's one-electron atomic antenna model \cite{Kuchiev.45.319} to an exact many-body description. For monitoring and following the dynamical correlations we employ measures from quantum information theory.\\
Applications to noble gases (He, Ne) and alkaline-earth atoms (Be, Mg) reveal striking differences. While the correlation measures in the noble gas atoms develop local maxima at each extremum of the driving laser field, surprisingly, in the alkaline earth atoms correlations feature local minima. By analyzing the matrix elements of the collision operator, we can attribute these dynamical changes in electronic correlations to the shape of the natural orbitals involved. At the present laser intensities, the natural orbitals approximately retain their original s- or p-wave character when they are distorted along the laser field. While in He and Ne the highest occupied natural orbital (HONO) and the lowest unoccupied natural orbital (LUNO) have the same parity, in Be and Mg the HONO has an (approximate) s-wave character while the LUNO has an (approximate) p-wave character.\\
To pinpoint the influence of correlations on HHG we decompose the exact evolution of the 1RDM in terms of the natural orbitals and that of their occupation numbers. While the time variation of the occupation numbers is exclusively due to two-particle correlations represented by the collision operator, the evolution of the natural orbitals is governed by a combination of the strong-field driving, mean-field effects and two-particle correlations. We have shown that the HHG spectra, in particular the high-energy plateau, are strongly and unambiguously influenced by two-particle correlations and their effect on the evolution of the natural orbitals, whereas the correlation-driven variation of the occupation numbers has only a minor impact on HHG. The time-dependence of two-particle correlations itself can be conveniently monitored by the occupation numbers and functions thereof such as the entropy. Their impact on the HHG mostly manifests themselves through their influence on the natural orbitals.\\
The present approach provides the foundation for future 
investigations of correlation effects in larger and more complex atoms such as Xe as well as extended systems. In Xe, the interplay between the outermost p- and d-shells leads to dramatic enhancements of the HHG yield in the spectral region of the giant dipole resonance (GDR). Understanding the correlation dynamics near the GDR under strong-field driving is still an open problem \cite{shiner_probing_2011, pabst_strong-field_2013, romanov_study_2021}. Even larger atoms may show similar effects and have the potential to dramatically increase the yield in the ultraviolet or extreme ultraviolet region promising technological applications. Systems with the number of electrons of Xe ($N=54$) or larger currently cannot be described within MCTDHF or even extensions thereof such as time-dependent complete active space self-consistent field methods \cite{sato_time-dependent_2013, sato_time-dependent_2016} due to the prohibitive exponential increase of the configuration space. Therefore, we plan to employ the polynomially scaling time-dependent two-particle reduced density matrix method \cite{lackner_propagating_2015, lackner_high-harmonic_2017, donsa_nonequilibrium_2023, Pescoller_projective_2025}, which is capable of accurately describing two-particle correlations, for studies of larger systems.
\section*{Acknowledgments}
This research was funded by the Austrian Science Fund (FWF) grant P 35539-N, as well as by the FWF 10.55776/COE1. 
Support of K.B. via the International Max Planck Research School of Advanced Photon Science is acknowledged. This research was supported in part by a Grant-in-Aid for Scientific Research (Grant No. JP22H05025, JP24H00427, and JP25K01688) from the Ministry of Education, Culture, Sports, Science and Technology (MEXT) of Japan. This research was also partially supported by the MEXT Quantum Leap Flagship Program, (Grant No. JPMXS0118067246) and JST COI-NEXT, (Grant No. JPMJPF2221). Calculations were performed on the Vienna Scientific Cluster (VSC4). \\

\appendix
\section{Numerical parameters for the simulations}\label{app:numerics}
For He ($N=Z=2$) we used $M=5$ orbitals with $l_\text{max} = 71$ expanded on a radial grid with $r_\text{max} = 200$ on $100$ elements within a finite elements discrete variable representation (FEDVR) \cite{rescigno_numerical_2000} with $20$ grid points in each element. An absorbing boundary of the shape $\text{cos}^4$ is applied, starting at $80\%$ of $r_\text{max}$. The ground state, obtained by imaginary time propagation, is propagated in velocity gauge using an exponential Runge-Kutta \cite{auzinger_efficient_2021} method with $20000$ time steps per optical cycle. Convergence of the results for these parameters has been verified.\\
For Ne ($N=Z=10$) with $M=9$ orbitals all parameters are kept as above but convergence is already reached for $l_\text{max} = 47$ and $50$ FEDVR elements with $20$ grid points in each element.\\ 
For Be ($N=Z=4$) with $M=5$ the radial grid was expanded to $r_\text{max} = 300$ with $75$ elements to reach convergence. All other parameters are the same as for He.
For Mg ($N=Z=12$) with $M=13$ the radial grid was expanded to $r_\text{max} = 280$ with $70$ elements and $l_\text{max} = 71$. For convergence $30000$ time steps per optical cycle are used. For the $l$ expansion of the electron-electron interaction (see \cite{sato_time-dependent_2016}) the same $l_\text{max}$ as for the expansion of the orbitals is used. All TDHF and TDDFT calculations use the same parameters as the MCTDHF calculations for the same atom.
The exchange and correlation functional is taken from the LibXC  \cite{lehtola_recent_2018} library. In the calculations shown PBE \cite{perdew_generalized_1997} functionals are used, but also the functinals LB94 \cite{gaiduk_construction_2011} and VWN \cite{vosko_accurate_1980} were checked, and only a weak dependence of the HHG spectrum on the functional used is observed.\\
For the calculation of the ionization probability (Eq.~\ref{eq:single_ion}), the ionization radius is set to $R_I = 20$ for all simulations. The time-dependent ionization probabilities are shifted in time to account for the time it takes the electron to propagate to $r = 20$, similar to the shift described in \cite{sato_time-dependent_2016}. Including this shift the maximua in $P_I$ coincide with the maximum field amplitude.
\section*{References}
\bibliography{references}

\begin{thebibliography}{80}%
\makeatletter
\providecommand \@ifxundefined [1]{%
 \@ifx{#1\undefined}
}%
\providecommand \@ifnum [1]{%
 \ifnum #1\expandafter \@firstoftwo
 \else \expandafter \@secondoftwo
 \fi
}%
\providecommand \@ifx [1]{%
 \ifx #1\expandafter \@firstoftwo
 \else \expandafter \@secondoftwo
 \fi
}%
\providecommand \natexlab [1]{#1}%
\providecommand \enquote  [1]{``#1''}%
\providecommand \bibnamefont  [1]{#1}%
\providecommand \bibfnamefont [1]{#1}%
\providecommand \citenamefont [1]{#1}%
\providecommand \href@noop [0]{\@secondoftwo}%
\providecommand \href [0]{\begingroup \@sanitize@url \@href}%
\providecommand \@href[1]{\@@startlink{#1}\@@href}%
\providecommand \@@href[1]{\endgroup#1\@@endlink}%
\providecommand \@sanitize@url [0]{\catcode `\\12\catcode `\$12\catcode `\&12\catcode `\#12\catcode `\^12\catcode `\_12\catcode `\%12\relax}%
\providecommand \@@startlink[1]{}%
\providecommand \@@endlink[0]{}%
\providecommand \url  [0]{\begingroup\@sanitize@url \@url }%
\providecommand \@url [1]{\endgroup\@href {#1}{\urlprefix }}%
\providecommand \urlprefix  [0]{URL }%
\providecommand \Eprint [0]{\href }%
\providecommand \doibase [0]{https://doi.org/}%
\providecommand \selectlanguage [0]{\@gobble}%
\providecommand \bibinfo  [0]{\@secondoftwo}%
\providecommand \bibfield  [0]{\@secondoftwo}%
\providecommand \translation [1]{[#1]}%
\providecommand \BibitemOpen [0]{}%
\providecommand \bibitemStop [0]{}%
\providecommand \bibitemNoStop [0]{.\EOS\space}%
\providecommand \EOS [0]{\spacefactor3000\relax}%
\providecommand \BibitemShut  [1]{\csname bibitem#1\endcsname}%
\let\auto@bib@innerbib\@empty
\bibitem [{\citenamefont {McPherson}\ \emph {et~al.}(1987)\citenamefont {McPherson}, \citenamefont {Gibson}, \citenamefont {Jara}, \citenamefont {Johann}, \citenamefont {Luk}, \citenamefont {McIntyre}, \citenamefont {Boyer},\ and\ \citenamefont {Rhodes}}]{mcpherson_studies_1987}%
  \BibitemOpen
  \bibfield  {author} {\bibinfo {author} {\bibfnamefont {A.}~\bibnamefont {McPherson}}, \bibinfo {author} {\bibfnamefont {G.}~\bibnamefont {Gibson}}, \bibinfo {author} {\bibfnamefont {H.}~\bibnamefont {Jara}}, \bibinfo {author} {\bibfnamefont {U.}~\bibnamefont {Johann}}, \bibinfo {author} {\bibfnamefont {T.~S.}\ \bibnamefont {Luk}}, \bibinfo {author} {\bibfnamefont {I.~A.}\ \bibnamefont {McIntyre}}, \bibinfo {author} {\bibfnamefont {K.}~\bibnamefont {Boyer}},\ and\ \bibinfo {author} {\bibfnamefont {C.~K.}\ \bibnamefont {Rhodes}},\ }\bibfield  {title} {\bibinfo {title} {Studies of multiphoton production of vacuum-ultraviolet radiation in the rare gases},\ }\href {https://doi.org/10.1364/JOSAB.4.000595} {\bibfield  {journal} {\bibinfo  {journal} {JOSA B}\ }\textbf {\bibinfo {volume} {4}},\ \bibinfo {pages} {595} (\bibinfo {year} {1987})},\ \bibinfo {note} {publisher: Optica Publishing Group}\BibitemShut {NoStop}%
\bibitem [{\citenamefont {Ferray}\ \emph {et~al.}(1988)\citenamefont {Ferray}, \citenamefont {L'Huillier}, \citenamefont {Li}, \citenamefont {Lompre}, \citenamefont {Mainfray},\ and\ \citenamefont {Manus}}]{ferray_multiple-harmonic_1988}%
  \BibitemOpen
  \bibfield  {author} {\bibinfo {author} {\bibfnamefont {M.}~\bibnamefont {Ferray}}, \bibinfo {author} {\bibfnamefont {A.}~\bibnamefont {L'Huillier}}, \bibinfo {author} {\bibfnamefont {X.~F.}\ \bibnamefont {Li}}, \bibinfo {author} {\bibfnamefont {L.~A.}\ \bibnamefont {Lompre}}, \bibinfo {author} {\bibfnamefont {G.}~\bibnamefont {Mainfray}},\ and\ \bibinfo {author} {\bibfnamefont {C.}~\bibnamefont {Manus}},\ }\bibfield  {title} {\bibinfo {title} {Multiple-harmonic conversion of 1064 nm radiation in rare gases},\ }\href {https://doi.org/10.1088/0953-4075/21/3/001} {\bibfield  {journal} {\bibinfo  {journal} {Journal of Physics B: Atomic, Molecular and Optical Physics}\ }\textbf {\bibinfo {volume} {21}},\ \bibinfo {pages} {L31} (\bibinfo {year} {1988})}\BibitemShut {NoStop}%
\bibitem [{\citenamefont {Schins}\ \emph {et~al.}(1994)\citenamefont {Schins}, \citenamefont {Breger}, \citenamefont {Agostini}, \citenamefont {Constantinescu}, \citenamefont {Muller}, \citenamefont {Grillon}, \citenamefont {Antonetti},\ and\ \citenamefont {Mysyrowicz}}]{schins_observation_1994}%
  \BibitemOpen
  \bibfield  {author} {\bibinfo {author} {\bibfnamefont {J.~M.}\ \bibnamefont {Schins}}, \bibinfo {author} {\bibfnamefont {P.}~\bibnamefont {Breger}}, \bibinfo {author} {\bibfnamefont {P.}~\bibnamefont {Agostini}}, \bibinfo {author} {\bibfnamefont {R.~C.}\ \bibnamefont {Constantinescu}}, \bibinfo {author} {\bibfnamefont {H.~G.}\ \bibnamefont {Muller}}, \bibinfo {author} {\bibfnamefont {G.}~\bibnamefont {Grillon}}, \bibinfo {author} {\bibfnamefont {A.}~\bibnamefont {Antonetti}},\ and\ \bibinfo {author} {\bibfnamefont {A.}~\bibnamefont {Mysyrowicz}},\ }\bibfield  {title} {\bibinfo {title} {Observation of {Laser}-{Assisted} {Auger} {Decay} in {Argon}},\ }\href {https://doi.org/10.1103/PhysRevLett.73.2180} {\bibfield  {journal} {\bibinfo  {journal} {Physical Review Letters}\ }\textbf {\bibinfo {volume} {73}},\ \bibinfo {pages} {2180} (\bibinfo {year} {1994})}\BibitemShut {NoStop}%
\bibitem [{\citenamefont {Véniard}\ \emph {et~al.}(1996)\citenamefont {Véniard}, \citenamefont {Taïeb},\ and\ \citenamefont {Maquet}}]{veniard_phase_1996}%
  \BibitemOpen
  \bibfield  {author} {\bibinfo {author} {\bibfnamefont {V.}~\bibnamefont {Véniard}}, \bibinfo {author} {\bibfnamefont {R.}~\bibnamefont {Taïeb}},\ and\ \bibinfo {author} {\bibfnamefont {A.}~\bibnamefont {Maquet}},\ }\bibfield  {title} {\bibinfo {title} {Phase dependence of ( \textit{{N}} +1)-color ( \textit{{N}} {\textgreater}1) ir-uv photoionization of atoms with higher harmonics},\ }\href {https://doi.org/10.1103/PhysRevA.54.721} {\bibfield  {journal} {\bibinfo  {journal} {Physical Review A}\ }\textbf {\bibinfo {volume} {54}},\ \bibinfo {pages} {721} (\bibinfo {year} {1996})}\BibitemShut {NoStop}%
\bibitem [{\citenamefont {Paul}\ \emph {et~al.}(2001)\citenamefont {Paul}, \citenamefont {Toma}, \citenamefont {Breger}, \citenamefont {Mullot}, \citenamefont {Augé}, \citenamefont {Balcou}, \citenamefont {Muller},\ and\ \citenamefont {Agostini}}]{paul_observation_2001}%
  \BibitemOpen
  \bibfield  {author} {\bibinfo {author} {\bibfnamefont {P.~M.}\ \bibnamefont {Paul}}, \bibinfo {author} {\bibfnamefont {E.~S.}\ \bibnamefont {Toma}}, \bibinfo {author} {\bibfnamefont {P.}~\bibnamefont {Breger}}, \bibinfo {author} {\bibfnamefont {G.}~\bibnamefont {Mullot}}, \bibinfo {author} {\bibfnamefont {F.}~\bibnamefont {Augé}}, \bibinfo {author} {\bibfnamefont {P.}~\bibnamefont {Balcou}}, \bibinfo {author} {\bibfnamefont {H.~G.}\ \bibnamefont {Muller}},\ and\ \bibinfo {author} {\bibfnamefont {P.}~\bibnamefont {Agostini}},\ }\bibfield  {title} {\bibinfo {title} {Observation of a {Train} of {Attosecond} {Pulses} from {High} {Harmonic} {Generation}},\ }\href {https://doi.org/10.1126/science.1059413} {\bibfield  {journal} {\bibinfo  {journal} {Science}\ }\textbf {\bibinfo {volume} {292}},\ \bibinfo {pages} {1689} (\bibinfo {year} {2001})}\BibitemShut {NoStop}%
\bibitem [{\citenamefont {Aseyev}\ \emph {et~al.}(2003)\citenamefont {Aseyev}, \citenamefont {Ni}, \citenamefont {Frasinski}, \citenamefont {Muller},\ and\ \citenamefont {Vrakking}}]{aseyev_attosecond_2003}%
  \BibitemOpen
  \bibfield  {author} {\bibinfo {author} {\bibfnamefont {S.~A.}\ \bibnamefont {Aseyev}}, \bibinfo {author} {\bibfnamefont {Y.}~\bibnamefont {Ni}}, \bibinfo {author} {\bibfnamefont {L.~J.}\ \bibnamefont {Frasinski}}, \bibinfo {author} {\bibfnamefont {H.~G.}\ \bibnamefont {Muller}},\ and\ \bibinfo {author} {\bibfnamefont {M.~J.~J.}\ \bibnamefont {Vrakking}},\ }\bibfield  {title} {\bibinfo {title} {Attosecond {Angle}-{Resolved} {Photoelectron} {Spectroscopy}},\ }\href {https://doi.org/10.1103/PhysRevLett.91.223902} {\bibfield  {journal} {\bibinfo  {journal} {Physical Review Letters}\ }\textbf {\bibinfo {volume} {91}},\ \bibinfo {pages} {223902} (\bibinfo {year} {2003})}\BibitemShut {NoStop}%
\bibitem [{\citenamefont {López-Martens}\ \emph {et~al.}(2005)\citenamefont {López-Martens}, \citenamefont {Varjú}, \citenamefont {Johnsson}, \citenamefont {Mauritsson}, \citenamefont {Mairesse}, \citenamefont {Salières}, \citenamefont {Gaarde}, \citenamefont {Schafer}, \citenamefont {Persson}, \citenamefont {Svanberg}, \citenamefont {Wahlström},\ and\ \citenamefont {L’Huillier}}]{lopez-martens_amplitude_2005}%
  \BibitemOpen
  \bibfield  {author} {\bibinfo {author} {\bibfnamefont {R.}~\bibnamefont {López-Martens}}, \bibinfo {author} {\bibfnamefont {K.}~\bibnamefont {Varjú}}, \bibinfo {author} {\bibfnamefont {P.}~\bibnamefont {Johnsson}}, \bibinfo {author} {\bibfnamefont {J.}~\bibnamefont {Mauritsson}}, \bibinfo {author} {\bibfnamefont {Y.}~\bibnamefont {Mairesse}}, \bibinfo {author} {\bibfnamefont {P.}~\bibnamefont {Salières}}, \bibinfo {author} {\bibfnamefont {M.~B.}\ \bibnamefont {Gaarde}}, \bibinfo {author} {\bibfnamefont {K.~J.}\ \bibnamefont {Schafer}}, \bibinfo {author} {\bibfnamefont {A.}~\bibnamefont {Persson}}, \bibinfo {author} {\bibfnamefont {S.}~\bibnamefont {Svanberg}}, \bibinfo {author} {\bibfnamefont {C.-G.}\ \bibnamefont {Wahlström}},\ and\ \bibinfo {author} {\bibfnamefont {A.}~\bibnamefont {L’Huillier}},\ }\bibfield  {title} {\bibinfo {title} {Amplitude and {Phase} {Control} of {Attosecond} {Light} {Pulses}},\ }\href {https://doi.org/10.1103/PhysRevLett.94.033001} {\bibfield  {journal} {\bibinfo  {journal}
  {Physical Review Letters}\ }\textbf {\bibinfo {volume} {94}},\ \bibinfo {pages} {033001} (\bibinfo {year} {2005})}\BibitemShut {NoStop}%
\bibitem [{\citenamefont {Nisoli}\ \emph {et~al.}(1997)\citenamefont {Nisoli}, \citenamefont {De~Silvestri}, \citenamefont {Svelto}, \citenamefont {Szipöcs}, \citenamefont {Ferencz}, \citenamefont {Spielmann}, \citenamefont {Sartania},\ and\ \citenamefont {Krausz}}]{nisoli_compression_1997}%
  \BibitemOpen
  \bibfield  {author} {\bibinfo {author} {\bibfnamefont {M.}~\bibnamefont {Nisoli}}, \bibinfo {author} {\bibfnamefont {S.}~\bibnamefont {De~Silvestri}}, \bibinfo {author} {\bibfnamefont {O.}~\bibnamefont {Svelto}}, \bibinfo {author} {\bibfnamefont {R.}~\bibnamefont {Szipöcs}}, \bibinfo {author} {\bibfnamefont {K.}~\bibnamefont {Ferencz}}, \bibinfo {author} {\bibfnamefont {C.}~\bibnamefont {Spielmann}}, \bibinfo {author} {\bibfnamefont {S.}~\bibnamefont {Sartania}},\ and\ \bibinfo {author} {\bibfnamefont {F.}~\bibnamefont {Krausz}},\ }\bibfield  {title} {\bibinfo {title} {Compression of high-energy laser pulses below 5 fs},\ }\href {https://doi.org/10.1364/OL.22.000522} {\bibfield  {journal} {\bibinfo  {journal} {Optics Letters}\ }\textbf {\bibinfo {volume} {22}},\ \bibinfo {pages} {522} (\bibinfo {year} {1997})}\BibitemShut {NoStop}%
\bibitem [{\citenamefont {Drescher}\ \emph {et~al.}(2002)\citenamefont {Drescher}, \citenamefont {Hentschel}, \citenamefont {Kienberger}, \citenamefont {Uiberacker}, \citenamefont {Yakovlev}, \citenamefont {Scrinzi}, \citenamefont {Westerwalbesloh}, \citenamefont {Kleineberg}, \citenamefont {Heinzmann},\ and\ \citenamefont {Krausz}}]{drescher_time-resolved_2002}%
  \BibitemOpen
  \bibfield  {author} {\bibinfo {author} {\bibfnamefont {M.}~\bibnamefont {Drescher}}, \bibinfo {author} {\bibfnamefont {M.}~\bibnamefont {Hentschel}}, \bibinfo {author} {\bibfnamefont {R.}~\bibnamefont {Kienberger}}, \bibinfo {author} {\bibfnamefont {M.}~\bibnamefont {Uiberacker}}, \bibinfo {author} {\bibfnamefont {V.}~\bibnamefont {Yakovlev}}, \bibinfo {author} {\bibfnamefont {A.}~\bibnamefont {Scrinzi}}, \bibinfo {author} {\bibfnamefont {T.}~\bibnamefont {Westerwalbesloh}}, \bibinfo {author} {\bibfnamefont {U.}~\bibnamefont {Kleineberg}}, \bibinfo {author} {\bibfnamefont {U.}~\bibnamefont {Heinzmann}},\ and\ \bibinfo {author} {\bibfnamefont {F.}~\bibnamefont {Krausz}},\ }\bibfield  {title} {\bibinfo {title} {Time-resolved atomic inner-shell spectroscopy},\ }\href {https://doi.org/10.1038/nature01143} {\bibfield  {journal} {\bibinfo  {journal} {Nature}\ }\textbf {\bibinfo {volume} {419}},\ \bibinfo {pages} {803} (\bibinfo {year} {2002})}\BibitemShut {NoStop}%
\bibitem [{\citenamefont {Kienberger}\ \emph {et~al.}(2004)\citenamefont {Kienberger}, \citenamefont {Goulielmakis}, \citenamefont {Uiberacker}, \citenamefont {Baltuska}, \citenamefont {Yakovlev}, \citenamefont {Bammer}, \citenamefont {Scrinzi}, \citenamefont {Westerwalbesloh}, \citenamefont {Kleineberg}, \citenamefont {Heinzmann}, \citenamefont {Drescher},\ and\ \citenamefont {Krausz}}]{kienberger_atomic_2004}%
  \BibitemOpen
  \bibfield  {author} {\bibinfo {author} {\bibfnamefont {R.}~\bibnamefont {Kienberger}}, \bibinfo {author} {\bibfnamefont {E.}~\bibnamefont {Goulielmakis}}, \bibinfo {author} {\bibfnamefont {M.}~\bibnamefont {Uiberacker}}, \bibinfo {author} {\bibfnamefont {A.}~\bibnamefont {Baltuska}}, \bibinfo {author} {\bibfnamefont {V.}~\bibnamefont {Yakovlev}}, \bibinfo {author} {\bibfnamefont {F.}~\bibnamefont {Bammer}}, \bibinfo {author} {\bibfnamefont {A.}~\bibnamefont {Scrinzi}}, \bibinfo {author} {\bibfnamefont {T.}~\bibnamefont {Westerwalbesloh}}, \bibinfo {author} {\bibfnamefont {U.}~\bibnamefont {Kleineberg}}, \bibinfo {author} {\bibfnamefont {U.}~\bibnamefont {Heinzmann}}, \bibinfo {author} {\bibfnamefont {M.}~\bibnamefont {Drescher}},\ and\ \bibinfo {author} {\bibfnamefont {F.}~\bibnamefont {Krausz}},\ }\bibfield  {title} {\bibinfo {title} {Atomic transient recorder},\ }\href {https://doi.org/10.1038/nature02277} {\bibfield  {journal} {\bibinfo  {journal} {Nature}\ }\textbf {\bibinfo {volume} {427}},\ \bibinfo
  {pages} {817} (\bibinfo {year} {2004})}\BibitemShut {NoStop}%
\bibitem [{\citenamefont {Santra}\ \emph {et~al.}(2011)\citenamefont {Santra}, \citenamefont {Yakovlev}, \citenamefont {Pfeifer},\ and\ \citenamefont {Loh}}]{santra_theory_2011}%
  \BibitemOpen
  \bibfield  {author} {\bibinfo {author} {\bibfnamefont {R.}~\bibnamefont {Santra}}, \bibinfo {author} {\bibfnamefont {V.~S.}\ \bibnamefont {Yakovlev}}, \bibinfo {author} {\bibfnamefont {T.}~\bibnamefont {Pfeifer}},\ and\ \bibinfo {author} {\bibfnamefont {Z.-H.}\ \bibnamefont {Loh}},\ }\bibfield  {title} {\bibinfo {title} {Theory of attosecond transient absorption spectroscopy of strong-field-generated ions},\ }\href {https://doi.org/10.1103/PhysRevA.83.033405} {\bibfield  {journal} {\bibinfo  {journal} {Physical Review A}\ }\textbf {\bibinfo {volume} {83}},\ \bibinfo {pages} {033405} (\bibinfo {year} {2011})}\BibitemShut {NoStop}%
\bibitem [{\citenamefont {Chen}\ \emph {et~al.}(2012)\citenamefont {Chen}, \citenamefont {Bell}, \citenamefont {Beck}, \citenamefont {Mashiko}, \citenamefont {Wu}, \citenamefont {Pfeiffer}, \citenamefont {Gaarde}, \citenamefont {Neumark}, \citenamefont {Leone},\ and\ \citenamefont {Schafer}}]{chen_light-induced_2012}%
  \BibitemOpen
  \bibfield  {author} {\bibinfo {author} {\bibfnamefont {S.}~\bibnamefont {Chen}}, \bibinfo {author} {\bibfnamefont {M.~J.}\ \bibnamefont {Bell}}, \bibinfo {author} {\bibfnamefont {A.~R.}\ \bibnamefont {Beck}}, \bibinfo {author} {\bibfnamefont {H.}~\bibnamefont {Mashiko}}, \bibinfo {author} {\bibfnamefont {M.}~\bibnamefont {Wu}}, \bibinfo {author} {\bibfnamefont {A.~N.}\ \bibnamefont {Pfeiffer}}, \bibinfo {author} {\bibfnamefont {M.~B.}\ \bibnamefont {Gaarde}}, \bibinfo {author} {\bibfnamefont {D.~M.}\ \bibnamefont {Neumark}}, \bibinfo {author} {\bibfnamefont {S.~R.}\ \bibnamefont {Leone}},\ and\ \bibinfo {author} {\bibfnamefont {K.~J.}\ \bibnamefont {Schafer}},\ }\bibfield  {title} {\bibinfo {title} {Light-induced states in attosecond transient absorption spectra of laser-dressed helium},\ }\href {https://doi.org/10.1103/PhysRevA.86.063408} {\bibfield  {journal} {\bibinfo  {journal} {Physical Review A}\ }\textbf {\bibinfo {volume} {86}},\ \bibinfo {pages} {063408} (\bibinfo {year} {2012})}\BibitemShut
  {NoStop}%
\bibitem [{\citenamefont {Spielmann}\ \emph {et~al.}(1997)\citenamefont {Spielmann}, \citenamefont {Burnett}, \citenamefont {Sartania}, \citenamefont {Koppitsch}, \citenamefont {Schnürer}, \citenamefont {Kan}, \citenamefont {Lenzner}, \citenamefont {Wobrauschek},\ and\ \citenamefont {Krausz}}]{spielmann_generation_1997}%
  \BibitemOpen
  \bibfield  {author} {\bibinfo {author} {\bibfnamefont {C.}~\bibnamefont {Spielmann}}, \bibinfo {author} {\bibfnamefont {N.~H.}\ \bibnamefont {Burnett}}, \bibinfo {author} {\bibfnamefont {S.}~\bibnamefont {Sartania}}, \bibinfo {author} {\bibfnamefont {R.}~\bibnamefont {Koppitsch}}, \bibinfo {author} {\bibfnamefont {M.}~\bibnamefont {Schnürer}}, \bibinfo {author} {\bibfnamefont {C.}~\bibnamefont {Kan}}, \bibinfo {author} {\bibfnamefont {M.}~\bibnamefont {Lenzner}}, \bibinfo {author} {\bibfnamefont {P.}~\bibnamefont {Wobrauschek}},\ and\ \bibinfo {author} {\bibfnamefont {F.}~\bibnamefont {Krausz}},\ }\bibfield  {title} {\bibinfo {title} {Generation of {Coherent} {X}-rays in the {Water} {Window} {Using} 5-{Femtosecond} {Laser} {Pulses}},\ }\href {https://doi.org/10.1126/science.278.5338.661} {\bibfield  {journal} {\bibinfo  {journal} {Science}\ }\textbf {\bibinfo {volume} {278}},\ \bibinfo {pages} {661} (\bibinfo {year} {1997})}\BibitemShut {NoStop}%
\bibitem [{\citenamefont {Popmintchev}\ \emph {et~al.}(2012)\citenamefont {Popmintchev}, \citenamefont {Chen}, \citenamefont {Popmintchev}, \citenamefont {Arpin}, \citenamefont {Brown}, \citenamefont {Ališauskas}, \citenamefont {Andriukaitis}, \citenamefont {Balčiunas}, \citenamefont {Mücke}, \citenamefont {Pugzlys}, \citenamefont {Baltuška}, \citenamefont {Shim}, \citenamefont {Schrauth}, \citenamefont {Gaeta}, \citenamefont {Hernández-García}, \citenamefont {Plaja}, \citenamefont {Becker}, \citenamefont {Jaron-Becker}, \citenamefont {Murnane},\ and\ \citenamefont {Kapteyn}}]{popmintchev_bright_2012}%
  \BibitemOpen
  \bibfield  {author} {\bibinfo {author} {\bibfnamefont {T.}~\bibnamefont {Popmintchev}}, \bibinfo {author} {\bibfnamefont {M.-C.}\ \bibnamefont {Chen}}, \bibinfo {author} {\bibfnamefont {D.}~\bibnamefont {Popmintchev}}, \bibinfo {author} {\bibfnamefont {P.}~\bibnamefont {Arpin}}, \bibinfo {author} {\bibfnamefont {S.}~\bibnamefont {Brown}}, \bibinfo {author} {\bibfnamefont {S.}~\bibnamefont {Ališauskas}}, \bibinfo {author} {\bibfnamefont {G.}~\bibnamefont {Andriukaitis}}, \bibinfo {author} {\bibfnamefont {T.}~\bibnamefont {Balčiunas}}, \bibinfo {author} {\bibfnamefont {O.~D.}\ \bibnamefont {Mücke}}, \bibinfo {author} {\bibfnamefont {A.}~\bibnamefont {Pugzlys}}, \bibinfo {author} {\bibfnamefont {A.}~\bibnamefont {Baltuška}}, \bibinfo {author} {\bibfnamefont {B.}~\bibnamefont {Shim}}, \bibinfo {author} {\bibfnamefont {S.~E.}\ \bibnamefont {Schrauth}}, \bibinfo {author} {\bibfnamefont {A.}~\bibnamefont {Gaeta}}, \bibinfo {author} {\bibfnamefont {C.}~\bibnamefont {Hernández-García}}, \bibinfo {author}
  {\bibfnamefont {L.}~\bibnamefont {Plaja}}, \bibinfo {author} {\bibfnamefont {A.}~\bibnamefont {Becker}}, \bibinfo {author} {\bibfnamefont {A.}~\bibnamefont {Jaron-Becker}}, \bibinfo {author} {\bibfnamefont {M.~M.}\ \bibnamefont {Murnane}},\ and\ \bibinfo {author} {\bibfnamefont {H.~C.}\ \bibnamefont {Kapteyn}},\ }\bibfield  {title} {\bibinfo {title} {Bright {Coherent} {Ultrahigh} {Harmonics} in the {keV} {X}-ray {Regime} from {Mid}-{Infrared} {Femtosecond} {Lasers}},\ }\href {https://doi.org/10.1126/science.1218497} {\bibfield  {journal} {\bibinfo  {journal} {Science}\ }\textbf {\bibinfo {volume} {336}},\ \bibinfo {pages} {1287} (\bibinfo {year} {2012})}\BibitemShut {NoStop}%
\bibitem [{\citenamefont {Hädrich}\ \emph {et~al.}(2015)\citenamefont {Hädrich}, \citenamefont {Krebs}, \citenamefont {Hoffmann}, \citenamefont {Klenke}, \citenamefont {Rothhardt}, \citenamefont {Limpert},\ and\ \citenamefont {Tünnermann}}]{hadrich_exploring_2015}%
  \BibitemOpen
  \bibfield  {author} {\bibinfo {author} {\bibfnamefont {S.}~\bibnamefont {Hädrich}}, \bibinfo {author} {\bibfnamefont {M.}~\bibnamefont {Krebs}}, \bibinfo {author} {\bibfnamefont {A.}~\bibnamefont {Hoffmann}}, \bibinfo {author} {\bibfnamefont {A.}~\bibnamefont {Klenke}}, \bibinfo {author} {\bibfnamefont {J.}~\bibnamefont {Rothhardt}}, \bibinfo {author} {\bibfnamefont {J.}~\bibnamefont {Limpert}},\ and\ \bibinfo {author} {\bibfnamefont {A.}~\bibnamefont {Tünnermann}},\ }\bibfield  {title} {\bibinfo {title} {Exploring new avenues in high repetition rate table-top coherent extreme ultraviolet sources},\ }\href {https://doi.org/10.1038/lsa.2015.93} {\bibfield  {journal} {\bibinfo  {journal} {Light: Science \& Applications}\ }\textbf {\bibinfo {volume} {4}},\ \bibinfo {pages} {e320} (\bibinfo {year} {2015})}\BibitemShut {NoStop}%
\bibitem [{\citenamefont {Naranjo-Montoya}\ \emph {et~al.}(2024)\citenamefont {Naranjo-Montoya}, \citenamefont {Bridger}, \citenamefont {Bhar}, \citenamefont {Kalkhoff}, \citenamefont {Schleberger}, \citenamefont {Wende}, \citenamefont {Tarasevitch},\ and\ \citenamefont {Bovensiepen}}]{naranjo-montoya_table-top_2024}%
  \BibitemOpen
  \bibfield  {author} {\bibinfo {author} {\bibfnamefont {O.~A.}\ \bibnamefont {Naranjo-Montoya}}, \bibinfo {author} {\bibfnamefont {M.}~\bibnamefont {Bridger}}, \bibinfo {author} {\bibfnamefont {R.}~\bibnamefont {Bhar}}, \bibinfo {author} {\bibfnamefont {L.}~\bibnamefont {Kalkhoff}}, \bibinfo {author} {\bibfnamefont {M.}~\bibnamefont {Schleberger}}, \bibinfo {author} {\bibfnamefont {H.}~\bibnamefont {Wende}}, \bibinfo {author} {\bibfnamefont {A.}~\bibnamefont {Tarasevitch}},\ and\ \bibinfo {author} {\bibfnamefont {U.}~\bibnamefont {Bovensiepen}},\ }\bibfield  {title} {\bibinfo {title} {Table-top source for x-ray absorption spectroscopy with photon energies up to 350 {eV}},\ }\href {https://doi.org/10.1063/5.0219921} {\bibfield  {journal} {\bibinfo  {journal} {Review of Scientific Instruments}\ }\textbf {\bibinfo {volume} {95}},\ \bibinfo {pages} {103001} (\bibinfo {year} {2024})}\BibitemShut {NoStop}%
\bibitem [{\citenamefont {Brabec}\ and\ \citenamefont {Krausz}(2000)}]{brabec_intense_2000}%
  \BibitemOpen
  \bibfield  {author} {\bibinfo {author} {\bibfnamefont {T.}~\bibnamefont {Brabec}}\ and\ \bibinfo {author} {\bibfnamefont {F.}~\bibnamefont {Krausz}},\ }\bibfield  {title} {\bibinfo {title} {Intense few-cycle laser fields: {Frontiers} of nonlinear optics},\ }\href {https://doi.org/10.1103/RevModPhys.72.545} {\bibfield  {journal} {\bibinfo  {journal} {Reviews of Modern Physics}\ }\textbf {\bibinfo {volume} {72}},\ \bibinfo {pages} {545} (\bibinfo {year} {2000})}\BibitemShut {NoStop}%
\bibitem [{\citenamefont {Krausz}\ and\ \citenamefont {Ivanov}(2009)}]{krausz_attosecond_2009}%
  \BibitemOpen
  \bibfield  {author} {\bibinfo {author} {\bibfnamefont {F.}~\bibnamefont {Krausz}}\ and\ \bibinfo {author} {\bibfnamefont {M.}~\bibnamefont {Ivanov}},\ }\bibfield  {title} {\bibinfo {title} {Attosecond physics},\ }\href {https://doi.org/10.1103/RevModPhys.81.163} {\bibfield  {journal} {\bibinfo  {journal} {Reviews of Modern Physics}\ }\textbf {\bibinfo {volume} {81}},\ \bibinfo {pages} {163} (\bibinfo {year} {2009})}\BibitemShut {NoStop}%
\bibitem [{\citenamefont {Kuchiev}(1987)}]{Kuchiev.45.319}%
  \BibitemOpen
  \bibfield  {author} {\bibinfo {author} {\bibfnamefont {M.~Y.}\ \bibnamefont {Kuchiev}},\ }\bibfield  {title} {\bibinfo {title} {Atomic antenna},\ }\href {http://jetpletters.ru/ps/0/article_18763.shtml} {\bibfield  {journal} {\bibinfo  {journal} {JETP letters}\ }\textbf {\bibinfo {volume} {45}},\ \bibinfo {pages} {404} (\bibinfo {year} {1987})}\BibitemShut {NoStop}%
\bibitem [{\citenamefont {Corkum}(1993)}]{corkum_plasma_1993}%
  \BibitemOpen
  \bibfield  {author} {\bibinfo {author} {\bibfnamefont {P.~B.}\ \bibnamefont {Corkum}},\ }\bibfield  {title} {{\selectlanguage {english}\bibinfo {title} {Plasma perspective on strong field multiphoton ionization}},\ }\href {https://doi.org/10.1103/PhysRevLett.71.1994} {\bibfield  {journal} {\bibinfo  {journal} {Physical Review Letters}\ }\textbf {\bibinfo {volume} {71}},\ \bibinfo {pages} {1994} (\bibinfo {year} {1993})}\BibitemShut {NoStop}%
\bibitem [{\citenamefont {Lewenstein}\ \emph {et~al.}(1994)\citenamefont {Lewenstein}, \citenamefont {Balcou}, \citenamefont {Ivanov}, \citenamefont {L’Huillier},\ and\ \citenamefont {Corkum}}]{lewenstein_theory_1994}%
  \BibitemOpen
  \bibfield  {author} {\bibinfo {author} {\bibfnamefont {M.}~\bibnamefont {Lewenstein}}, \bibinfo {author} {\bibfnamefont {P.}~\bibnamefont {Balcou}}, \bibinfo {author} {\bibfnamefont {M.~Y.}\ \bibnamefont {Ivanov}}, \bibinfo {author} {\bibfnamefont {A.}~\bibnamefont {L’Huillier}},\ and\ \bibinfo {author} {\bibfnamefont {P.~B.}\ \bibnamefont {Corkum}},\ }\bibfield  {title} {\bibinfo {title} {Theory of high-harmonic generation by low-frequency laser fields},\ }\href {https://doi.org/10.1103/PhysRevA.49.2117} {\bibfield  {journal} {\bibinfo  {journal} {Physical Review A}\ }\textbf {\bibinfo {volume} {49}},\ \bibinfo {pages} {2117} (\bibinfo {year} {1994})}\BibitemShut {NoStop}%
\bibitem [{\citenamefont {Shiner}\ \emph {et~al.}(2011)\citenamefont {Shiner}, \citenamefont {Schmidt}, \citenamefont {Trallero-Herrero}, \citenamefont {Wörner}, \citenamefont {Patchkovskii}, \citenamefont {Corkum}, \citenamefont {Kieffer}, \citenamefont {Légaré},\ and\ \citenamefont {Villeneuve}}]{shiner_probing_2011}%
  \BibitemOpen
  \bibfield  {author} {\bibinfo {author} {\bibfnamefont {A.~D.}\ \bibnamefont {Shiner}}, \bibinfo {author} {\bibfnamefont {B.~E.}\ \bibnamefont {Schmidt}}, \bibinfo {author} {\bibfnamefont {C.}~\bibnamefont {Trallero-Herrero}}, \bibinfo {author} {\bibfnamefont {H.~J.}\ \bibnamefont {Wörner}}, \bibinfo {author} {\bibfnamefont {S.}~\bibnamefont {Patchkovskii}}, \bibinfo {author} {\bibfnamefont {P.~B.}\ \bibnamefont {Corkum}}, \bibinfo {author} {\bibfnamefont {J.-C.}\ \bibnamefont {Kieffer}}, \bibinfo {author} {\bibfnamefont {F.}~\bibnamefont {Légaré}},\ and\ \bibinfo {author} {\bibfnamefont {D.~M.}\ \bibnamefont {Villeneuve}},\ }\bibfield  {title} {\bibinfo {title} {Probing collective multi-electron dynamics in xenon with high-harmonic spectroscopy},\ }\href {https://doi.org/10.1038/nphys1940} {\bibfield  {journal} {\bibinfo  {journal} {Nature Physics}\ }\textbf {\bibinfo {volume} {7}},\ \bibinfo {pages} {464} (\bibinfo {year} {2011})}\BibitemShut {NoStop}%
\bibitem [{\citenamefont {Pabst}\ and\ \citenamefont {Santra}(2013)}]{pabst_strong-field_2013}%
  \BibitemOpen
  \bibfield  {author} {\bibinfo {author} {\bibfnamefont {S.}~\bibnamefont {Pabst}}\ and\ \bibinfo {author} {\bibfnamefont {R.}~\bibnamefont {Santra}},\ }\bibfield  {title} {\bibinfo {title} {Strong-{Field} {Many}-{Body} {Physics} and the {Giant} {Enhancement} in the {High}-{Harmonic} {Spectrum} of {Xenon}},\ }\href {https://doi.org/10.1103/PhysRevLett.111.233005} {\bibfield  {journal} {\bibinfo  {journal} {Physical Review Letters}\ }\textbf {\bibinfo {volume} {111}},\ \bibinfo {pages} {233005} (\bibinfo {year} {2013})}\BibitemShut {NoStop}%
\bibitem [{\citenamefont {Ganeev}\ \emph {et~al.}(2006)\citenamefont {Ganeev}, \citenamefont {Suzuki}, \citenamefont {Baba}, \citenamefont {Kuroda},\ and\ \citenamefont {Ozaki}}]{ganeev_strong_2006}%
  \BibitemOpen
  \bibfield  {author} {\bibinfo {author} {\bibfnamefont {R.~A.}\ \bibnamefont {Ganeev}}, \bibinfo {author} {\bibfnamefont {M.}~\bibnamefont {Suzuki}}, \bibinfo {author} {\bibfnamefont {M.}~\bibnamefont {Baba}}, \bibinfo {author} {\bibfnamefont {H.}~\bibnamefont {Kuroda}},\ and\ \bibinfo {author} {\bibfnamefont {T.}~\bibnamefont {Ozaki}},\ }\bibfield  {title} {\bibinfo {title} {Strong resonance enhancement of a single harmonic generated in the extreme ultraviolet range},\ }\href {https://doi.org/10.1364/OL.31.001699} {\bibfield  {journal} {\bibinfo  {journal} {Optics Letters}\ }\textbf {\bibinfo {volume} {31}},\ \bibinfo {pages} {1699} (\bibinfo {year} {2006})}\BibitemShut {NoStop}%
\bibitem [{\citenamefont {Ganeev}\ \emph {et~al.}(2007)\citenamefont {Ganeev}, \citenamefont {Naik}, \citenamefont {Singhal}, \citenamefont {Chakera},\ and\ \citenamefont {Gupta}}]{ganeev_strong_2007}%
  \BibitemOpen
  \bibfield  {author} {\bibinfo {author} {\bibfnamefont {R.~A.}\ \bibnamefont {Ganeev}}, \bibinfo {author} {\bibfnamefont {P.~A.}\ \bibnamefont {Naik}}, \bibinfo {author} {\bibfnamefont {H.}~\bibnamefont {Singhal}}, \bibinfo {author} {\bibfnamefont {J.~A.}\ \bibnamefont {Chakera}},\ and\ \bibinfo {author} {\bibfnamefont {P.~D.}\ \bibnamefont {Gupta}},\ }\bibfield  {title} {\bibinfo {title} {Strong enhancement and extinction of single harmonic intensity in the mid- and end-plateau regions of the high harmonics generated in weakly excited laser plasmas},\ }\href {https://doi.org/10.1364/OL.32.000065} {\bibfield  {journal} {\bibinfo  {journal} {Optics Letters}\ }\textbf {\bibinfo {volume} {32}},\ \bibinfo {pages} {65} (\bibinfo {year} {2007})}\BibitemShut {NoStop}%
\bibitem [{\citenamefont {Ganeev}\ \emph {et~al.}(2012)\citenamefont {Ganeev}, \citenamefont {Strelkov}, \citenamefont {Hutchison}, \citenamefont {Zaïr}, \citenamefont {Kilbane}, \citenamefont {Khokhlova},\ and\ \citenamefont {Marangos}}]{ganeev_experimental_2012}%
  \BibitemOpen
  \bibfield  {author} {\bibinfo {author} {\bibfnamefont {R.~A.}\ \bibnamefont {Ganeev}}, \bibinfo {author} {\bibfnamefont {V.~V.}\ \bibnamefont {Strelkov}}, \bibinfo {author} {\bibfnamefont {C.}~\bibnamefont {Hutchison}}, \bibinfo {author} {\bibfnamefont {A.}~\bibnamefont {Zaïr}}, \bibinfo {author} {\bibfnamefont {D.}~\bibnamefont {Kilbane}}, \bibinfo {author} {\bibfnamefont {M.~A.}\ \bibnamefont {Khokhlova}},\ and\ \bibinfo {author} {\bibfnamefont {J.~P.}\ \bibnamefont {Marangos}},\ }\bibfield  {title} {\bibinfo {title} {Experimental and theoretical studies of two-color-pump resonance-induced enhancement of odd and even harmonics from a tin plasma},\ }\href {https://doi.org/10.1103/PhysRevA.85.023832} {\bibfield  {journal} {\bibinfo  {journal} {Physical Review A}\ }\textbf {\bibinfo {volume} {85}},\ \bibinfo {pages} {023832} (\bibinfo {year} {2012})}\BibitemShut {NoStop}%
\bibitem [{\citenamefont {Haessler}\ \emph {et~al.}(2013)\citenamefont {Haessler}, \citenamefont {Strelkov}, \citenamefont {Elouga~Bom}, \citenamefont {Khokhlova}, \citenamefont {Gobert}, \citenamefont {Hergott}, \citenamefont {Lepetit}, \citenamefont {Perdrix}, \citenamefont {Ozaki},\ and\ \citenamefont {Salières}}]{haessler_phase_2013}%
  \BibitemOpen
  \bibfield  {author} {\bibinfo {author} {\bibfnamefont {S.}~\bibnamefont {Haessler}}, \bibinfo {author} {\bibfnamefont {V.}~\bibnamefont {Strelkov}}, \bibinfo {author} {\bibfnamefont {L.~B.}\ \bibnamefont {Elouga~Bom}}, \bibinfo {author} {\bibfnamefont {M.}~\bibnamefont {Khokhlova}}, \bibinfo {author} {\bibfnamefont {O.}~\bibnamefont {Gobert}}, \bibinfo {author} {\bibfnamefont {J.-F.}\ \bibnamefont {Hergott}}, \bibinfo {author} {\bibfnamefont {F.}~\bibnamefont {Lepetit}}, \bibinfo {author} {\bibfnamefont {M.}~\bibnamefont {Perdrix}}, \bibinfo {author} {\bibfnamefont {T.}~\bibnamefont {Ozaki}},\ and\ \bibinfo {author} {\bibfnamefont {P.}~\bibnamefont {Salières}},\ }\bibfield  {title} {\bibinfo {title} {Phase distortions of attosecond pulses produced by resonance-enhanced high harmonic generation},\ }\href {https://doi.org/10.1088/1367-2630/15/1/013051} {\bibfield  {journal} {\bibinfo  {journal} {New Journal of Physics}\ }\textbf {\bibinfo {volume} {15}},\ \bibinfo {pages} {013051} (\bibinfo {year}
  {2013})}\BibitemShut {NoStop}%
\bibitem [{\citenamefont {Wahyutama}\ \emph {et~al.}(2019)\citenamefont {Wahyutama}, \citenamefont {Sato},\ and\ \citenamefont {Ishikawa}}]{wahyutama_time-dependent_2019}%
  \BibitemOpen
  \bibfield  {author} {\bibinfo {author} {\bibfnamefont {I.~S.}\ \bibnamefont {Wahyutama}}, \bibinfo {author} {\bibfnamefont {T.}~\bibnamefont {Sato}},\ and\ \bibinfo {author} {\bibfnamefont {K.~L.}\ \bibnamefont {Ishikawa}},\ }\bibfield  {title} {\bibinfo {title} {Time-dependent multiconfiguration self-consistent-field study on resonantly enhanced high-order harmonic generation from transition-metal elements},\ }\href {https://doi.org/10.1103/PhysRevA.99.063420} {\bibfield  {journal} {\bibinfo  {journal} {Physical Review A}\ }\textbf {\bibinfo {volume} {99}},\ \bibinfo {pages} {063420} (\bibinfo {year} {2019})}\BibitemShut {NoStop}%
\bibitem [{\citenamefont {Bray}\ \emph {et~al.}(2020)\citenamefont {Bray}, \citenamefont {Freeman}, \citenamefont {Naseem}, \citenamefont {Dolmatov},\ and\ \citenamefont {Kheifets}}]{bray_correlation-enhanced_2020}%
  \BibitemOpen
  \bibfield  {author} {\bibinfo {author} {\bibfnamefont {A.~W.}\ \bibnamefont {Bray}}, \bibinfo {author} {\bibfnamefont {D.}~\bibnamefont {Freeman}}, \bibinfo {author} {\bibfnamefont {F.}~\bibnamefont {Naseem}}, \bibinfo {author} {\bibfnamefont {V.~K.}\ \bibnamefont {Dolmatov}},\ and\ \bibinfo {author} {\bibfnamefont {A.~S.}\ \bibnamefont {Kheifets}},\ }\bibfield  {title} {\bibinfo {title} {Correlation-enhanced high-order-harmonic-generation spectra of {Mn} and {Mn} +},\ }\href {https://doi.org/10.1103/PhysRevA.101.053415} {\bibfield  {journal} {\bibinfo  {journal} {Physical Review A}\ }\textbf {\bibinfo {volume} {101}},\ \bibinfo {pages} {053415} (\bibinfo {year} {2020})}\BibitemShut {NoStop}%
\bibitem [{\citenamefont {Artemyev}\ \emph {et~al.}(2017)\citenamefont {Artemyev}, \citenamefont {Cederbaum},\ and\ \citenamefont {Demekhin}}]{artemyev_impact_2017}%
  \BibitemOpen
  \bibfield  {author} {\bibinfo {author} {\bibfnamefont {A.~N.}\ \bibnamefont {Artemyev}}, \bibinfo {author} {\bibfnamefont {L.~S.}\ \bibnamefont {Cederbaum}},\ and\ \bibinfo {author} {\bibfnamefont {P.~V.}\ \bibnamefont {Demekhin}},\ }\bibfield  {title} {\bibinfo {title} {Impact of two-electron dynamics and correlations on high-order-harmonic generation in {He}},\ }\href {https://doi.org/10.1103/PhysRevA.95.033402} {\bibfield  {journal} {\bibinfo  {journal} {Physical Review A}\ }\textbf {\bibinfo {volume} {95}},\ \bibinfo {pages} {033402} (\bibinfo {year} {2017})}\BibitemShut {NoStop}%
\bibitem [{\citenamefont {De~Las~Heras}\ \emph {et~al.}(2020)\citenamefont {De~Las~Heras}, \citenamefont {Hernández-García},\ and\ \citenamefont {Plaja}}]{de_las_heras_spectral_2020}%
  \BibitemOpen
  \bibfield  {author} {\bibinfo {author} {\bibfnamefont {A.}~\bibnamefont {De~Las~Heras}}, \bibinfo {author} {\bibfnamefont {C.}~\bibnamefont {Hernández-García}},\ and\ \bibinfo {author} {\bibfnamefont {L.}~\bibnamefont {Plaja}},\ }\bibfield  {title} {{\selectlanguage {english}\bibinfo {title} {Spectral signature of back reaction in correlated electron dynamics in intense electromagnetic fields}},\ }\href {https://doi.org/10.1103/PhysRevResearch.2.033047} {\bibfield  {journal} {\bibinfo  {journal} {Physical Review Research}\ }\textbf {\bibinfo {volume} {2}},\ \bibinfo {pages} {033047} (\bibinfo {year} {2020})}\BibitemShut {NoStop}%
\bibitem [{\citenamefont {Tikhomirov}\ \emph {et~al.}(2017)\citenamefont {Tikhomirov}, \citenamefont {Sato},\ and\ \citenamefont {Ishikawa}}]{tikhomirov_high-harmonic_2017}%
  \BibitemOpen
  \bibfield  {author} {\bibinfo {author} {\bibfnamefont {I.}~\bibnamefont {Tikhomirov}}, \bibinfo {author} {\bibfnamefont {T.}~\bibnamefont {Sato}},\ and\ \bibinfo {author} {\bibfnamefont {K.~L.}\ \bibnamefont {Ishikawa}},\ }\bibfield  {title} {{\selectlanguage {english}\bibinfo {title} {High-{Harmonic} {Generation} {Enhanced} by {Dynamical} {Electron} {Correlation}}},\ }\href {https://doi.org/10.1103/PhysRevLett.118.203202} {\bibfield  {journal} {\bibinfo  {journal} {Physical Review Letters}\ }\textbf {\bibinfo {volume} {118}},\ \bibinfo {pages} {203202} (\bibinfo {year} {2017})}\BibitemShut {NoStop}%
\bibitem [{\citenamefont {Li}\ \emph {et~al.}(2019)\citenamefont {Li}, \citenamefont {Sato},\ and\ \citenamefont {Ishikawa}}]{li_high-order_2019}%
  \BibitemOpen
  \bibfield  {author} {\bibinfo {author} {\bibfnamefont {Y.}~\bibnamefont {Li}}, \bibinfo {author} {\bibfnamefont {T.}~\bibnamefont {Sato}},\ and\ \bibinfo {author} {\bibfnamefont {K.~L.}\ \bibnamefont {Ishikawa}},\ }\bibfield  {title} {\bibinfo {title} {High-order harmonic generation enhanced by laser-induced electron recollision},\ }\href {https://doi.org/10.1103/PhysRevA.99.043401} {\bibfield  {journal} {\bibinfo  {journal} {Physical Review A}\ }\textbf {\bibinfo {volume} {99}},\ \bibinfo {pages} {043401} (\bibinfo {year} {2019})}\BibitemShut {NoStop}%
\bibitem [{\citenamefont {Neufeld}\ and\ \citenamefont {Cohen}(2020)}]{neufeld_probing_2020}%
  \BibitemOpen
  \bibfield  {author} {\bibinfo {author} {\bibfnamefont {O.}~\bibnamefont {Neufeld}}\ and\ \bibinfo {author} {\bibfnamefont {O.}~\bibnamefont {Cohen}},\ }\bibfield  {title} {{\selectlanguage {english}\bibinfo {title} {Probing ultrafast electron correlations in high harmonic generation}},\ }\href {https://doi.org/10.1103/PhysRevResearch.2.033037} {\bibfield  {journal} {\bibinfo  {journal} {Physical Review Research}\ }\textbf {\bibinfo {volume} {2}},\ \bibinfo {pages} {033037} (\bibinfo {year} {2020})}\BibitemShut {NoStop}%
\bibitem [{\citenamefont {Ghimire}\ \emph {et~al.}(2011)\citenamefont {Ghimire}, \citenamefont {DiChiara}, \citenamefont {Sistrunk}, \citenamefont {Agostini}, \citenamefont {DiMauro},\ and\ \citenamefont {Reis}}]{ghimire_observation_2011}%
  \BibitemOpen
  \bibfield  {author} {\bibinfo {author} {\bibfnamefont {S.}~\bibnamefont {Ghimire}}, \bibinfo {author} {\bibfnamefont {A.~D.}\ \bibnamefont {DiChiara}}, \bibinfo {author} {\bibfnamefont {E.}~\bibnamefont {Sistrunk}}, \bibinfo {author} {\bibfnamefont {P.}~\bibnamefont {Agostini}}, \bibinfo {author} {\bibfnamefont {L.~F.}\ \bibnamefont {DiMauro}},\ and\ \bibinfo {author} {\bibfnamefont {D.~A.}\ \bibnamefont {Reis}},\ }\bibfield  {title} {\bibinfo {title} {Observation of high-order harmonic generation in a bulk crystal},\ }\href {https://doi.org/10.1038/nphys1847} {\bibfield  {journal} {\bibinfo  {journal} {Nature Physics}\ }\textbf {\bibinfo {volume} {7}},\ \bibinfo {pages} {138} (\bibinfo {year} {2011})},\ \bibinfo {note} {publisher: Nature Publishing Group}\BibitemShut {NoStop}%
\bibitem [{\citenamefont {Ghimire}\ \emph {et~al.}(2014)\citenamefont {Ghimire}, \citenamefont {Ndabashimiye}, \citenamefont {DiChiara}, \citenamefont {Sistrunk}, \citenamefont {Stockman}, \citenamefont {Agostini}, \citenamefont {DiMauro},\ and\ \citenamefont {Reis}}]{ghimire_strong-field_2014}%
  \BibitemOpen
  \bibfield  {author} {\bibinfo {author} {\bibfnamefont {S.}~\bibnamefont {Ghimire}}, \bibinfo {author} {\bibfnamefont {G.}~\bibnamefont {Ndabashimiye}}, \bibinfo {author} {\bibfnamefont {A.~D.}\ \bibnamefont {DiChiara}}, \bibinfo {author} {\bibfnamefont {E.}~\bibnamefont {Sistrunk}}, \bibinfo {author} {\bibfnamefont {M.~I.}\ \bibnamefont {Stockman}}, \bibinfo {author} {\bibfnamefont {P.}~\bibnamefont {Agostini}}, \bibinfo {author} {\bibfnamefont {L.~F.}\ \bibnamefont {DiMauro}},\ and\ \bibinfo {author} {\bibfnamefont {D.~A.}\ \bibnamefont {Reis}},\ }\bibfield  {title} {\bibinfo {title} {Strong-field and attosecond physics in solids},\ }\href {https://doi.org/10.1088/0953-4075/47/20/204030} {\bibfield  {journal} {\bibinfo  {journal} {Journal of Physics B: Atomic, Molecular and Optical Physics}\ }\textbf {\bibinfo {volume} {47}},\ \bibinfo {pages} {204030} (\bibinfo {year} {2014})},\ \bibinfo {note} {publisher: IOP Publishing}\BibitemShut {NoStop}%
\bibitem [{\citenamefont {Luu}\ \emph {et~al.}(2015)\citenamefont {Luu}, \citenamefont {Garg}, \citenamefont {Kruchinin}, \citenamefont {Moulet}, \citenamefont {Hassan},\ and\ \citenamefont {Goulielmakis}}]{luu_extreme_2015}%
  \BibitemOpen
  \bibfield  {author} {\bibinfo {author} {\bibfnamefont {T.~T.}\ \bibnamefont {Luu}}, \bibinfo {author} {\bibfnamefont {M.}~\bibnamefont {Garg}}, \bibinfo {author} {\bibfnamefont {S.~Y.}\ \bibnamefont {Kruchinin}}, \bibinfo {author} {\bibfnamefont {A.}~\bibnamefont {Moulet}}, \bibinfo {author} {\bibfnamefont {M.~T.}\ \bibnamefont {Hassan}},\ and\ \bibinfo {author} {\bibfnamefont {E.}~\bibnamefont {Goulielmakis}},\ }\bibfield  {title} {\bibinfo {title} {Extreme ultraviolet high-harmonic spectroscopy of solids},\ }\href {https://doi.org/10.1038/nature14456} {\bibfield  {journal} {\bibinfo  {journal} {Nature}\ }\textbf {\bibinfo {volume} {521}},\ \bibinfo {pages} {498} (\bibinfo {year} {2015})},\ \bibinfo {note} {publisher: Nature Publishing Group}\BibitemShut {NoStop}%
\bibitem [{\citenamefont {Vampa}\ \emph {et~al.}(2015)\citenamefont {Vampa}, \citenamefont {Hammond}, \citenamefont {Thiré}, \citenamefont {Schmidt}, \citenamefont {Légaré}, \citenamefont {McDonald}, \citenamefont {Brabec}, \citenamefont {Klug},\ and\ \citenamefont {Corkum}}]{vampa_all-optical_2015}%
  \BibitemOpen
  \bibfield  {author} {\bibinfo {author} {\bibfnamefont {G.}~\bibnamefont {Vampa}}, \bibinfo {author} {\bibfnamefont {T.}~\bibnamefont {Hammond}}, \bibinfo {author} {\bibfnamefont {N.}~\bibnamefont {Thiré}}, \bibinfo {author} {\bibfnamefont {B.}~\bibnamefont {Schmidt}}, \bibinfo {author} {\bibfnamefont {F.}~\bibnamefont {Légaré}}, \bibinfo {author} {\bibfnamefont {C.}~\bibnamefont {McDonald}}, \bibinfo {author} {\bibfnamefont {T.}~\bibnamefont {Brabec}}, \bibinfo {author} {\bibfnamefont {D.}~\bibnamefont {Klug}},\ and\ \bibinfo {author} {\bibfnamefont {P.}~\bibnamefont {Corkum}},\ }\bibfield  {title} {\bibinfo {title} {All-{Optical} {Reconstruction} of {Crystal} {Band} {Structure}},\ }\href {https://doi.org/10.1103/PhysRevLett.115.193603} {\bibfield  {journal} {\bibinfo  {journal} {Physical Review Letters}\ }\textbf {\bibinfo {volume} {115}},\ \bibinfo {pages} {193603} (\bibinfo {year} {2015})},\ \bibinfo {note} {publisher: American Physical Society}\BibitemShut {NoStop}%
\bibitem [{\citenamefont {Floss}\ \emph {et~al.}(2018)\citenamefont {Floss}, \citenamefont {Lemell}, \citenamefont {Wachter}, \citenamefont {Smejkal}, \citenamefont {Sato}, \citenamefont {Tong}, \citenamefont {Yabana},\ and\ \citenamefont {Burgdörfer}}]{floss_ab_2018}%
  \BibitemOpen
  \bibfield  {author} {\bibinfo {author} {\bibfnamefont {I.}~\bibnamefont {Floss}}, \bibinfo {author} {\bibfnamefont {C.}~\bibnamefont {Lemell}}, \bibinfo {author} {\bibfnamefont {G.}~\bibnamefont {Wachter}}, \bibinfo {author} {\bibfnamefont {V.}~\bibnamefont {Smejkal}}, \bibinfo {author} {\bibfnamefont {S.~A.}\ \bibnamefont {Sato}}, \bibinfo {author} {\bibfnamefont {X.-M.}\ \bibnamefont {Tong}}, \bibinfo {author} {\bibfnamefont {K.}~\bibnamefont {Yabana}},\ and\ \bibinfo {author} {\bibfnamefont {J.}~\bibnamefont {Burgdörfer}},\ }\bibfield  {title} {\bibinfo {title} {Ab initio multiscale simulation of high-order harmonic generation in solids},\ }\href {https://doi.org/10.1103/PhysRevA.97.011401} {\bibfield  {journal} {\bibinfo  {journal} {Physical Review A}\ }\textbf {\bibinfo {volume} {97}},\ \bibinfo {pages} {011401} (\bibinfo {year} {2018})},\ \bibinfo {note} {publisher: American Physical Society}\BibitemShut {NoStop}%
\bibitem [{\citenamefont {Silva}\ \emph {et~al.}(2018)\citenamefont {Silva}, \citenamefont {Blinov}, \citenamefont {Rubtsov}, \citenamefont {Smirnova},\ and\ \citenamefont {Ivanov}}]{silva_high-harmonic_2018}%
  \BibitemOpen
  \bibfield  {author} {\bibinfo {author} {\bibfnamefont {R.~E.~F.}\ \bibnamefont {Silva}}, \bibinfo {author} {\bibfnamefont {I.~V.}\ \bibnamefont {Blinov}}, \bibinfo {author} {\bibfnamefont {A.~N.}\ \bibnamefont {Rubtsov}}, \bibinfo {author} {\bibfnamefont {O.}~\bibnamefont {Smirnova}},\ and\ \bibinfo {author} {\bibfnamefont {M.}~\bibnamefont {Ivanov}},\ }\bibfield  {title} {{\selectlanguage {english}\bibinfo {title} {High-harmonic spectroscopy of ultrafast many-body dynamics in strongly correlated systems}},\ }\href {https://doi.org/10.1038/s41566-018-0129-0} {\bibfield  {journal} {\bibinfo  {journal} {Nature Photonics}\ }\textbf {\bibinfo {volume} {12}},\ \bibinfo {pages} {266} (\bibinfo {year} {2018})}\BibitemShut {NoStop}%
\bibitem [{\citenamefont {Tancogne-Dejean}\ \emph {et~al.}(2018)\citenamefont {Tancogne-Dejean}, \citenamefont {Sentef},\ and\ \citenamefont {Rubio}}]{tancogne-dejean_ultrafast_2018}%
  \BibitemOpen
  \bibfield  {author} {\bibinfo {author} {\bibfnamefont {N.}~\bibnamefont {Tancogne-Dejean}}, \bibinfo {author} {\bibfnamefont {M.~A.}\ \bibnamefont {Sentef}},\ and\ \bibinfo {author} {\bibfnamefont {A.}~\bibnamefont {Rubio}},\ }\bibfield  {title} {{\selectlanguage {english}\bibinfo {title} {Ultrafast {Modification} of {Hubbard} {U} in a {Strongly} {Correlated} {Material}: \textit{{Ab} initio} {High}-{Harmonic} {Generation} in {NiO}}},\ }\href {https://doi.org/10.1103/PhysRevLett.121.097402} {\bibfield  {journal} {\bibinfo  {journal} {Physical Review Letters}\ }\textbf {\bibinfo {volume} {121}},\ \bibinfo {pages} {097402} (\bibinfo {year} {2018})}\BibitemShut {NoStop}%
\bibitem [{\citenamefont {Baykusheva}\ \emph {et~al.}(2022)\citenamefont {Baykusheva}, \citenamefont {Jang}, \citenamefont {Husain}, \citenamefont {Lee}, \citenamefont {TenHuisen}, \citenamefont {Zhou}, \citenamefont {Park}, \citenamefont {Kim}, \citenamefont {Kim}, \citenamefont {Kim}, \citenamefont {Kim}, \citenamefont {Park}, \citenamefont {Abbamonte}, \citenamefont {Kim}, \citenamefont {Gu}, \citenamefont {Wang},\ and\ \citenamefont {Mitrano}}]{baykusheva_ultrafast_2022}%
  \BibitemOpen
  \bibfield  {author} {\bibinfo {author} {\bibfnamefont {D.~R.}\ \bibnamefont {Baykusheva}}, \bibinfo {author} {\bibfnamefont {H.}~\bibnamefont {Jang}}, \bibinfo {author} {\bibfnamefont {A.~A.}\ \bibnamefont {Husain}}, \bibinfo {author} {\bibfnamefont {S.}~\bibnamefont {Lee}}, \bibinfo {author} {\bibfnamefont {S.~F.}\ \bibnamefont {TenHuisen}}, \bibinfo {author} {\bibfnamefont {P.}~\bibnamefont {Zhou}}, \bibinfo {author} {\bibfnamefont {S.}~\bibnamefont {Park}}, \bibinfo {author} {\bibfnamefont {H.}~\bibnamefont {Kim}}, \bibinfo {author} {\bibfnamefont {J.-K.}\ \bibnamefont {Kim}}, \bibinfo {author} {\bibfnamefont {H.-D.}\ \bibnamefont {Kim}}, \bibinfo {author} {\bibfnamefont {M.}~\bibnamefont {Kim}}, \bibinfo {author} {\bibfnamefont {S.-Y.}\ \bibnamefont {Park}}, \bibinfo {author} {\bibfnamefont {P.}~\bibnamefont {Abbamonte}}, \bibinfo {author} {\bibfnamefont {B.}~\bibnamefont {Kim}}, \bibinfo {author} {\bibfnamefont {G.}~\bibnamefont {Gu}}, \bibinfo {author} {\bibfnamefont {Y.}~\bibnamefont {Wang}},\ and\
  \bibinfo {author} {\bibfnamefont {M.}~\bibnamefont {Mitrano}},\ }\bibfield  {title} {\bibinfo {title} {Ultrafast {Renormalization} of the {On}-{Site} {Coulomb} {Repulsion} in a {Cuprate} {Superconductor}},\ }\href {https://doi.org/10.1103/PhysRevX.12.011013} {\bibfield  {journal} {\bibinfo  {journal} {Physical Review X}\ }\textbf {\bibinfo {volume} {12}},\ \bibinfo {pages} {011013} (\bibinfo {year} {2022})}\BibitemShut {NoStop}%
\bibitem [{\citenamefont {Caillat}\ \emph {et~al.}(2005)\citenamefont {Caillat}, \citenamefont {Zanghellini}, \citenamefont {Kitzler}, \citenamefont {Koch}, \citenamefont {Kreuzer},\ and\ \citenamefont {Scrinzi}}]{caillat_correlated_2005}%
  \BibitemOpen
  \bibfield  {author} {\bibinfo {author} {\bibfnamefont {J.}~\bibnamefont {Caillat}}, \bibinfo {author} {\bibfnamefont {J.}~\bibnamefont {Zanghellini}}, \bibinfo {author} {\bibfnamefont {M.}~\bibnamefont {Kitzler}}, \bibinfo {author} {\bibfnamefont {O.}~\bibnamefont {Koch}}, \bibinfo {author} {\bibfnamefont {W.}~\bibnamefont {Kreuzer}},\ and\ \bibinfo {author} {\bibfnamefont {A.}~\bibnamefont {Scrinzi}},\ }\bibfield  {title} {\bibinfo {title} {Correlated multielectron systems in strong laser fields: {A} multiconfiguration time-dependent {Hartree}-{Fock} approach},\ }\href {https://doi.org/10.1103/PhysRevA.71.012712} {\bibfield  {journal} {\bibinfo  {journal} {Physical Review A}\ }\textbf {\bibinfo {volume} {71}},\ \bibinfo {pages} {012712} (\bibinfo {year} {2005})}\BibitemShut {NoStop}%
\bibitem [{\citenamefont {Kato}\ and\ \citenamefont {Kono}(2004)}]{kato_time-dependent_2004}%
  \BibitemOpen
  \bibfield  {author} {\bibinfo {author} {\bibfnamefont {T.}~\bibnamefont {Kato}}\ and\ \bibinfo {author} {\bibfnamefont {H.}~\bibnamefont {Kono}},\ }\bibfield  {title} {\bibinfo {title} {Time-dependent multiconfiguration theory for electronic dynamics of molecules in an intense laser field},\ }\href {https://doi.org/10.1016/j.cplett.2004.05.106} {\bibfield  {journal} {\bibinfo  {journal} {Chemical Physics Letters}\ }\textbf {\bibinfo {volume} {392}},\ \bibinfo {pages} {533} (\bibinfo {year} {2004})}\BibitemShut {NoStop}%
\bibitem [{\citenamefont {Nest}\ \emph {et~al.}(2005)\citenamefont {Nest}, \citenamefont {Klamroth},\ and\ \citenamefont {Saalfrank}}]{nest_multiconfiguration_2005}%
  \BibitemOpen
  \bibfield  {author} {\bibinfo {author} {\bibfnamefont {M.}~\bibnamefont {Nest}}, \bibinfo {author} {\bibfnamefont {T.}~\bibnamefont {Klamroth}},\ and\ \bibinfo {author} {\bibfnamefont {P.}~\bibnamefont {Saalfrank}},\ }\bibfield  {title} {\bibinfo {title} {The multiconfiguration time-dependent {Hartree}–{Fock} method for quantum chemical calculations},\ }\href {https://doi.org/10.1063/1.1862243} {\bibfield  {journal} {\bibinfo  {journal} {The Journal of Chemical Physics}\ }\textbf {\bibinfo {volume} {122}},\ \bibinfo {pages} {124102} (\bibinfo {year} {2005})}\BibitemShut {NoStop}%
\bibitem [{\citenamefont {Haxton}\ \emph {et~al.}(2011)\citenamefont {Haxton}, \citenamefont {Lawler},\ and\ \citenamefont {McCurdy}}]{haxton_multiconfiguration_2011}%
  \BibitemOpen
  \bibfield  {author} {\bibinfo {author} {\bibfnamefont {D.~J.}\ \bibnamefont {Haxton}}, \bibinfo {author} {\bibfnamefont {K.~V.}\ \bibnamefont {Lawler}},\ and\ \bibinfo {author} {\bibfnamefont {C.~W.}\ \bibnamefont {McCurdy}},\ }\bibfield  {title} {\bibinfo {title} {Multiconfiguration time-dependent {Hartree}-{Fock} treatment of electronic and nuclear dynamics in diatomic molecules},\ }\href {https://doi.org/10.1103/PhysRevA.83.063416} {\bibfield  {journal} {\bibinfo  {journal} {Physical Review A}\ }\textbf {\bibinfo {volume} {83}},\ \bibinfo {pages} {063416} (\bibinfo {year} {2011})},\ \bibinfo {note} {publisher: American Physical Society}\BibitemShut {NoStop}%
\bibitem [{\citenamefont {Karabanov}\ \emph {et~al.}(2011)\citenamefont {Karabanov}, \citenamefont {Kuprov}, \citenamefont {Charnock}, \citenamefont {van~der Drift}, \citenamefont {Edwards},\ and\ \citenamefont {Köckenberger}}]{karabanov_accuracy_2011}%
  \BibitemOpen
  \bibfield  {author} {\bibinfo {author} {\bibfnamefont {A.}~\bibnamefont {Karabanov}}, \bibinfo {author} {\bibfnamefont {I.}~\bibnamefont {Kuprov}}, \bibinfo {author} {\bibfnamefont {G.~T.~P.}\ \bibnamefont {Charnock}}, \bibinfo {author} {\bibfnamefont {A.}~\bibnamefont {van~der Drift}}, \bibinfo {author} {\bibfnamefont {L.~J.}\ \bibnamefont {Edwards}},\ and\ \bibinfo {author} {\bibfnamefont {W.}~\bibnamefont {Köckenberger}},\ }\bibfield  {title} {\bibinfo {title} {On the accuracy of the state space restriction approximation for spin dynamics simulations},\ }\href {https://doi.org/10.1063/1.3624564} {\bibfield  {journal} {\bibinfo  {journal} {The Journal of Chemical Physics}\ }\textbf {\bibinfo {volume} {135}},\ \bibinfo {pages} {084106} (\bibinfo {year} {2011})}\BibitemShut {NoStop}%
\bibitem [{\citenamefont {Sato}\ \emph {et~al.}(2016)\citenamefont {Sato}, \citenamefont {Ishikawa}, \citenamefont {Březinová}, \citenamefont {Lackner}, \citenamefont {Nagele},\ and\ \citenamefont {Burgdörfer}}]{sato_time-dependent_2016}%
  \BibitemOpen
  \bibfield  {author} {\bibinfo {author} {\bibfnamefont {T.}~\bibnamefont {Sato}}, \bibinfo {author} {\bibfnamefont {K.~L.}\ \bibnamefont {Ishikawa}}, \bibinfo {author} {\bibfnamefont {I.}~\bibnamefont {Březinová}}, \bibinfo {author} {\bibfnamefont {F.}~\bibnamefont {Lackner}}, \bibinfo {author} {\bibfnamefont {S.}~\bibnamefont {Nagele}},\ and\ \bibinfo {author} {\bibfnamefont {J.}~\bibnamefont {Burgdörfer}},\ }\bibfield  {title} {\bibinfo {title} {Time-dependent complete-active-space self-consistent-field method for atoms: {Application} to high-order harmonic generation},\ }\href {https://doi.org/10.1103/PhysRevA.94.023405} {\bibfield  {journal} {\bibinfo  {journal} {Physical Review A}\ }\textbf {\bibinfo {volume} {94}},\ \bibinfo {pages} {023405} (\bibinfo {year} {2016})}\BibitemShut {NoStop}%
\bibitem [{\citenamefont {Kulander}\ \emph {et~al.}(1992)\citenamefont {Kulander}, \citenamefont {Schafer},\ and\ \citenamefont {Krause}}]{kulander_schafer_krause_1992}%
  \BibitemOpen
  \bibfield  {author} {\bibinfo {author} {\bibfnamefont {K.}~\bibnamefont {Kulander}}, \bibinfo {author} {\bibfnamefont {K.}~\bibnamefont {Schafer}},\ and\ \bibinfo {author} {\bibfnamefont {J.}~\bibnamefont {Krause}},\ }\href@noop {} {\emph {\bibinfo {title} {Time-dependent studies of multiphoton processes}}}\ (\bibinfo  {publisher} {Academic Press Inc.},\ \bibinfo {year} {1992})\ p.\ \bibinfo {pages} {p. 247–300.}\BibitemShut {Stop}%
\bibitem [{\citenamefont {Runge}\ and\ \citenamefont {Gross}(1984)}]{runge_density-functional_1984}%
  \BibitemOpen
  \bibfield  {author} {\bibinfo {author} {\bibfnamefont {E.}~\bibnamefont {Runge}}\ and\ \bibinfo {author} {\bibfnamefont {E.~K.~U.}\ \bibnamefont {Gross}},\ }\bibfield  {title} {\bibinfo {title} {Density-{Functional} {Theory} for {Time}-{Dependent} {Systems}},\ }\href {https://doi.org/10.1103/PhysRevLett.52.997} {\bibfield  {journal} {\bibinfo  {journal} {Physical Review Letters}\ }\textbf {\bibinfo {volume} {52}},\ \bibinfo {pages} {997} (\bibinfo {year} {1984})},\ \bibinfo {note} {publisher: American Physical Society}\BibitemShut {NoStop}%
\bibitem [{\citenamefont {Bogoljubov}(1985)}]{bogoljubov_lectures_1985}%
  \BibitemOpen
  \bibfield  {author} {\bibinfo {author} {\bibfnamefont {N.~N.}\ \bibnamefont {Bogoljubov}},\ }\href@noop {} {\emph {\bibinfo {title} {Lectures on quantum statistics: in two volumes. 1: {Quantum} statistics}}},\ \bibinfo {edition} {2nd}\ ed.\ (\bibinfo  {publisher} {Gordon and Breach},\ \bibinfo {address} {New York},\ \bibinfo {year} {1985})\BibitemShut {NoStop}%
\bibitem [{\citenamefont {Mazziotti}(1998)}]{mazziotti_approximate_1998}%
  \BibitemOpen
  \bibfield  {author} {\bibinfo {author} {\bibfnamefont {D.~A.}\ \bibnamefont {Mazziotti}},\ }\bibfield  {title} {\bibinfo {title} {Approximate solution for electron correlation through the use of {Schwinger} probes},\ }\href {https://doi.org/10.1016/S0009-2614(98)00470-9} {\bibfield  {journal} {\bibinfo  {journal} {Chemical Physics Letters}\ }\textbf {\bibinfo {volume} {289}},\ \bibinfo {pages} {419} (\bibinfo {year} {1998})}\BibitemShut {NoStop}%
\bibitem [{\citenamefont {Kutzelnigg}\ and\ \citenamefont {Mukherjee}(1999)}]{kutzelnigg_cumulant_1999}%
  \BibitemOpen
  \bibfield  {author} {\bibinfo {author} {\bibfnamefont {W.}~\bibnamefont {Kutzelnigg}}\ and\ \bibinfo {author} {\bibfnamefont {D.}~\bibnamefont {Mukherjee}},\ }\bibfield  {title} {\bibinfo {title} {Cumulant expansion of the reduced density matrices},\ }\href {https://doi.org/10.1063/1.478189} {\bibfield  {journal} {\bibinfo  {journal} {The Journal of Chemical Physics}\ }\textbf {\bibinfo {volume} {110}},\ \bibinfo {pages} {2800} (\bibinfo {year} {1999})}\BibitemShut {NoStop}%
\bibitem [{\citenamefont {Furche}(2001)}]{furche_density_2001}%
  \BibitemOpen
  \bibfield  {author} {\bibinfo {author} {\bibfnamefont {F.}~\bibnamefont {Furche}},\ }\bibfield  {title} {\bibinfo {title} {On the density matrix based approach to time-dependent density functional response theory},\ }\href {https://doi.org/10.1063/1.1353585} {\bibfield  {journal} {\bibinfo  {journal} {The Journal of Chemical Physics}\ }\textbf {\bibinfo {volume} {114}},\ \bibinfo {pages} {5982} (\bibinfo {year} {2001})},\ \bibinfo {note} {publisher: AIP Publishing}\BibitemShut {NoStop}%
\bibitem [{\citenamefont {Li}\ and\ \citenamefont {Ullrich}(2011)}]{li_time-dependent_2011}%
  \BibitemOpen
  \bibfield  {author} {\bibinfo {author} {\bibfnamefont {Y.}~\bibnamefont {Li}}\ and\ \bibinfo {author} {\bibfnamefont {C.}~\bibnamefont {Ullrich}},\ }\bibfield  {title} {\bibinfo {title} {Time-dependent transition density matrix},\ }\href {https://doi.org/10.1016/j.chemphys.2011.02.001} {\bibfield  {journal} {\bibinfo  {journal} {Chemical Physics}\ }\textbf {\bibinfo {volume} {391}},\ \bibinfo {pages} {157} (\bibinfo {year} {2011})},\ \bibinfo {note} {publisher: Elsevier BV}\BibitemShut {NoStop}%
\bibitem [{\citenamefont {Nielsen}\ and\ \citenamefont {Chuang}(2010)}]{nielsen_quantum_2010}%
  \BibitemOpen
  \bibfield  {author} {\bibinfo {author} {\bibfnamefont {M.~A.}\ \bibnamefont {Nielsen}}\ and\ \bibinfo {author} {\bibfnamefont {I.~L.}\ \bibnamefont {Chuang}},\ }\href@noop {} {\emph {\bibinfo {title} {Quantum computation and quantum information}}},\ \bibinfo {edition} {10th}\ ed.\ (\bibinfo  {publisher} {Cambridge university press},\ \bibinfo {address} {Cambridge},\ \bibinfo {year} {2010})\BibitemShut {NoStop}%
\bibitem [{\citenamefont {Bose}\ and\ \citenamefont {Vedral}(2000)}]{bose_mixedness_2000}%
  \BibitemOpen
  \bibfield  {author} {\bibinfo {author} {\bibfnamefont {S.}~\bibnamefont {Bose}}\ and\ \bibinfo {author} {\bibfnamefont {V.}~\bibnamefont {Vedral}},\ }\bibfield  {title} {\bibinfo {title} {Mixedness and teleportation},\ }\href {https://doi.org/10.1103/PhysRevA.61.040101} {\bibfield  {journal} {\bibinfo  {journal} {Physical Review A}\ }\textbf {\bibinfo {volume} {61}},\ \bibinfo {pages} {040101} (\bibinfo {year} {2000})}\BibitemShut {NoStop}%
\bibitem [{\citenamefont {Bourassin-Bouchet}\ \emph {et~al.}(2020)\citenamefont {Bourassin-Bouchet}, \citenamefont {Barreau}, \citenamefont {Gruson}, \citenamefont {Hergott}, \citenamefont {Quéré}, \citenamefont {Salières},\ and\ \citenamefont {Ruchon}}]{bourassin-bouchet_quantifying_2020}%
  \BibitemOpen
  \bibfield  {author} {\bibinfo {author} {\bibfnamefont {C.}~\bibnamefont {Bourassin-Bouchet}}, \bibinfo {author} {\bibfnamefont {L.}~\bibnamefont {Barreau}}, \bibinfo {author} {\bibfnamefont {V.}~\bibnamefont {Gruson}}, \bibinfo {author} {\bibfnamefont {J.-F.}\ \bibnamefont {Hergott}}, \bibinfo {author} {\bibfnamefont {F.}~\bibnamefont {Quéré}}, \bibinfo {author} {\bibfnamefont {P.}~\bibnamefont {Salières}},\ and\ \bibinfo {author} {\bibfnamefont {T.}~\bibnamefont {Ruchon}},\ }\bibfield  {title} {\bibinfo {title} {Quantifying {Decoherence} in {Attosecond} {Metrology}},\ }\href {https://doi.org/10.1103/PhysRevX.10.031048} {\bibfield  {journal} {\bibinfo  {journal} {Physical Review X}\ }\textbf {\bibinfo {volume} {10}},\ \bibinfo {pages} {031048} (\bibinfo {year} {2020})}\BibitemShut {NoStop}%
\bibitem [{\citenamefont {Vrakking}(2021)}]{vrakking_control_2021}%
  \BibitemOpen
  \bibfield  {author} {\bibinfo {author} {\bibfnamefont {M.~J.}\ \bibnamefont {Vrakking}},\ }\bibfield  {title} {{\selectlanguage {english}\bibinfo {title} {Control of {Attosecond} {Entanglement} and {Coherence}}},\ }\href {https://doi.org/10.1103/PhysRevLett.126.113203} {\bibfield  {journal} {\bibinfo  {journal} {Physical Review Letters}\ }\textbf {\bibinfo {volume} {126}},\ \bibinfo {pages} {113203} (\bibinfo {year} {2021})}\BibitemShut {NoStop}%
\bibitem [{\citenamefont {Koll}\ \emph {et~al.}(2022)\citenamefont {Koll}, \citenamefont {Maikowski}, \citenamefont {Drescher}, \citenamefont {Witting},\ and\ \citenamefont {Vrakking}}]{koll_experimental_2022}%
  \BibitemOpen
  \bibfield  {author} {\bibinfo {author} {\bibfnamefont {L.-M.}\ \bibnamefont {Koll}}, \bibinfo {author} {\bibfnamefont {L.}~\bibnamefont {Maikowski}}, \bibinfo {author} {\bibfnamefont {L.}~\bibnamefont {Drescher}}, \bibinfo {author} {\bibfnamefont {T.}~\bibnamefont {Witting}},\ and\ \bibinfo {author} {\bibfnamefont {M.~J.}\ \bibnamefont {Vrakking}},\ }\bibfield  {title} {{\selectlanguage {english}\bibinfo {title} {Experimental {Control} of {Quantum}-{Mechanical} {Entanglement} in an {Attosecond} {Pump}-{Probe} {Experiment}}},\ }\href {https://doi.org/10.1103/PhysRevLett.128.043201} {\bibfield  {journal} {\bibinfo  {journal} {Physical Review Letters}\ }\textbf {\bibinfo {volume} {128}},\ \bibinfo {pages} {043201} (\bibinfo {year} {2022})}\BibitemShut {NoStop}%
\bibitem [{\citenamefont {Nandi}\ \emph {et~al.}(2024)\citenamefont {Nandi}, \citenamefont {Stenquist}, \citenamefont {Papoulia}, \citenamefont {Olofsson}, \citenamefont {Badano}, \citenamefont {Bertolino}, \citenamefont {Busto}, \citenamefont {Callegari}, \citenamefont {Carlström}, \citenamefont {Danailov}, \citenamefont {Demekhin}, \citenamefont {Di~Fraia}, \citenamefont {Eng-Johnsson}, \citenamefont {Feifel}, \citenamefont {Gallician}, \citenamefont {Giannessi}, \citenamefont {Gisselbrecht}, \citenamefont {Manfredda}, \citenamefont {Meyer}, \citenamefont {Miron}, \citenamefont {Peschel}, \citenamefont {Plekan}, \citenamefont {Prince}, \citenamefont {Squibb}, \citenamefont {Zangrando}, \citenamefont {Zapata}, \citenamefont {Zhong},\ and\ \citenamefont {Dahlström}}]{nandi_generation_2024}%
  \BibitemOpen
  \bibfield  {author} {\bibinfo {author} {\bibfnamefont {S.}~\bibnamefont {Nandi}}, \bibinfo {author} {\bibfnamefont {A.}~\bibnamefont {Stenquist}}, \bibinfo {author} {\bibfnamefont {A.}~\bibnamefont {Papoulia}}, \bibinfo {author} {\bibfnamefont {E.}~\bibnamefont {Olofsson}}, \bibinfo {author} {\bibfnamefont {L.}~\bibnamefont {Badano}}, \bibinfo {author} {\bibfnamefont {M.}~\bibnamefont {Bertolino}}, \bibinfo {author} {\bibfnamefont {D.}~\bibnamefont {Busto}}, \bibinfo {author} {\bibfnamefont {C.}~\bibnamefont {Callegari}}, \bibinfo {author} {\bibfnamefont {S.}~\bibnamefont {Carlström}}, \bibinfo {author} {\bibfnamefont {M.~B.}\ \bibnamefont {Danailov}}, \bibinfo {author} {\bibfnamefont {P.~V.}\ \bibnamefont {Demekhin}}, \bibinfo {author} {\bibfnamefont {M.}~\bibnamefont {Di~Fraia}}, \bibinfo {author} {\bibfnamefont {P.}~\bibnamefont {Eng-Johnsson}}, \bibinfo {author} {\bibfnamefont {R.}~\bibnamefont {Feifel}}, \bibinfo {author} {\bibfnamefont {G.}~\bibnamefont {Gallician}}, \bibinfo {author} {\bibfnamefont
  {L.}~\bibnamefont {Giannessi}}, \bibinfo {author} {\bibfnamefont {M.}~\bibnamefont {Gisselbrecht}}, \bibinfo {author} {\bibfnamefont {M.}~\bibnamefont {Manfredda}}, \bibinfo {author} {\bibfnamefont {M.}~\bibnamefont {Meyer}}, \bibinfo {author} {\bibfnamefont {C.}~\bibnamefont {Miron}}, \bibinfo {author} {\bibfnamefont {J.}~\bibnamefont {Peschel}}, \bibinfo {author} {\bibfnamefont {O.}~\bibnamefont {Plekan}}, \bibinfo {author} {\bibfnamefont {K.~C.}\ \bibnamefont {Prince}}, \bibinfo {author} {\bibfnamefont {R.~J.}\ \bibnamefont {Squibb}}, \bibinfo {author} {\bibfnamefont {M.}~\bibnamefont {Zangrando}}, \bibinfo {author} {\bibfnamefont {F.}~\bibnamefont {Zapata}}, \bibinfo {author} {\bibfnamefont {S.}~\bibnamefont {Zhong}},\ and\ \bibinfo {author} {\bibfnamefont {J.~M.}\ \bibnamefont {Dahlström}},\ }\bibfield  {title} {\bibinfo {title} {Generation of entanglement using a short-wavelength seeded free-electron laser},\ }\href {https://doi.org/10.1126/sciadv.ado0668} {\bibfield  {journal} {\bibinfo  {journal}
  {Science Advances}\ }\textbf {\bibinfo {volume} {10}},\ \bibinfo {pages} {eado0668} (\bibinfo {year} {2024})}\BibitemShut {NoStop}%
\bibitem [{\citenamefont {Jiang}\ \emph {et~al.}(2024)\citenamefont {Jiang}, \citenamefont {Zhong}, \citenamefont {Fang}, \citenamefont {Donsa}, \citenamefont {Březinová}, \citenamefont {Peng},\ and\ \citenamefont {Burgdörfer}}]{jiang_time_2024}%
  \BibitemOpen
  \bibfield  {author} {\bibinfo {author} {\bibfnamefont {W.-C.}\ \bibnamefont {Jiang}}, \bibinfo {author} {\bibfnamefont {M.-C.}\ \bibnamefont {Zhong}}, \bibinfo {author} {\bibfnamefont {Y.-K.}\ \bibnamefont {Fang}}, \bibinfo {author} {\bibfnamefont {S.}~\bibnamefont {Donsa}}, \bibinfo {author} {\bibfnamefont {I.}~\bibnamefont {Březinová}}, \bibinfo {author} {\bibfnamefont {L.-Y.}\ \bibnamefont {Peng}},\ and\ \bibinfo {author} {\bibfnamefont {J.}~\bibnamefont {Burgdörfer}},\ }\bibfield  {title} {\bibinfo {title} {Time {Delays} as {Attosecond} {Probe} of {Interelectronic} {Coherence} and {Entanglement}},\ }\href {https://doi.org/10.1103/PhysRevLett.133.163201} {\bibfield  {journal} {\bibinfo  {journal} {Physical Review Letters}\ }\textbf {\bibinfo {volume} {133}},\ \bibinfo {pages} {163201} (\bibinfo {year} {2024})},\ \bibinfo {note} {publisher: American Physical Society}\BibitemShut {NoStop}%
\bibitem [{\citenamefont {Ishikawa}\ \emph {et~al.}(2023)\citenamefont {Ishikawa}, \citenamefont {Prince},\ and\ \citenamefont {Ueda}}]{ishikawa_control_2023}%
  \BibitemOpen
  \bibfield  {author} {\bibinfo {author} {\bibfnamefont {K.~L.}\ \bibnamefont {Ishikawa}}, \bibinfo {author} {\bibfnamefont {K.~C.}\ \bibnamefont {Prince}},\ and\ \bibinfo {author} {\bibfnamefont {K.}~\bibnamefont {Ueda}},\ }\bibfield  {title} {\bibinfo {title} {Control of {Ion}-{Photoelectron} {Entanglement} and {Coherence} {Via} {Rabi} {Oscillations}},\ }\href {https://doi.org/10.1021/acs.jpca.3c06781} {\bibfield  {journal} {\bibinfo  {journal} {The Journal of Physical Chemistry A}\ }\textbf {\bibinfo {volume} {127}},\ \bibinfo {pages} {10638} (\bibinfo {year} {2023})}\BibitemShut {NoStop}%
\bibitem [{\citenamefont {Shen}\ \emph {et~al.}(2025)\citenamefont {Shen}, \citenamefont {Mao}, \citenamefont {Zhang}, \citenamefont {Li}, \citenamefont {Sato}, \citenamefont {Ishikawa},\ and\ \citenamefont {He}}]{shen_coherent_2025}%
  \BibitemOpen
  \bibfield  {author} {\bibinfo {author} {\bibfnamefont {B.-R.}\ \bibnamefont {Shen}}, \bibinfo {author} {\bibfnamefont {Y.-J.}\ \bibnamefont {Mao}}, \bibinfo {author} {\bibfnamefont {Z.-H.}\ \bibnamefont {Zhang}}, \bibinfo {author} {\bibfnamefont {Y.}~\bibnamefont {Li}}, \bibinfo {author} {\bibfnamefont {T.}~\bibnamefont {Sato}}, \bibinfo {author} {\bibfnamefont {K.~L.}\ \bibnamefont {Ishikawa}},\ and\ \bibinfo {author} {\bibfnamefont {F.}~\bibnamefont {He}},\ }\bibfield  {title} {\bibinfo {title} {Coherent control of ion-photoelectron dynamics through {Rabi} oscillations: {An} \textit{ab initio} study},\ }\href {https://doi.org/10.1103/c338-8n6w} {\bibfield  {journal} {\bibinfo  {journal} {Physical Review A}\ }\textbf {\bibinfo {volume} {111}},\ \bibinfo {pages} {063113} (\bibinfo {year} {2025})}\BibitemShut {NoStop}%
\bibitem [{\citenamefont {Laurell}\ \emph {et~al.}(2022)\citenamefont {Laurell}, \citenamefont {Finkelstein-Shapiro}, \citenamefont {Dittel}, \citenamefont {Guo}, \citenamefont {Demjaha}, \citenamefont {Ammitzböll}, \citenamefont {Weissenbilder}, \citenamefont {Neoričić}, \citenamefont {Luo}, \citenamefont {Gisselbrecht}, \citenamefont {Arnold}, \citenamefont {Buchleitner}, \citenamefont {Pullerits}, \citenamefont {L'Huillier},\ and\ \citenamefont {Busto}}]{laurell_continuous-variable_2022}%
  \BibitemOpen
  \bibfield  {author} {\bibinfo {author} {\bibfnamefont {H.}~\bibnamefont {Laurell}}, \bibinfo {author} {\bibfnamefont {D.}~\bibnamefont {Finkelstein-Shapiro}}, \bibinfo {author} {\bibfnamefont {C.}~\bibnamefont {Dittel}}, \bibinfo {author} {\bibfnamefont {C.}~\bibnamefont {Guo}}, \bibinfo {author} {\bibfnamefont {R.}~\bibnamefont {Demjaha}}, \bibinfo {author} {\bibfnamefont {M.}~\bibnamefont {Ammitzböll}}, \bibinfo {author} {\bibfnamefont {R.}~\bibnamefont {Weissenbilder}}, \bibinfo {author} {\bibfnamefont {L.}~\bibnamefont {Neoričić}}, \bibinfo {author} {\bibfnamefont {S.}~\bibnamefont {Luo}}, \bibinfo {author} {\bibfnamefont {M.}~\bibnamefont {Gisselbrecht}}, \bibinfo {author} {\bibfnamefont {C.~L.}\ \bibnamefont {Arnold}}, \bibinfo {author} {\bibfnamefont {A.}~\bibnamefont {Buchleitner}}, \bibinfo {author} {\bibfnamefont {T.}~\bibnamefont {Pullerits}}, \bibinfo {author} {\bibfnamefont {A.}~\bibnamefont {L'Huillier}},\ and\ \bibinfo {author} {\bibfnamefont {D.}~\bibnamefont {Busto}},\ }\bibfield
  {title} {\bibinfo {title} {Continuous-variable quantum state tomography of photoelectrons},\ }\href {https://doi.org/10.1103/PhysRevResearch.4.033220} {\bibfield  {journal} {\bibinfo  {journal} {Physical Review Research}\ }\textbf {\bibinfo {volume} {4}},\ \bibinfo {pages} {033220} (\bibinfo {year} {2022})}\BibitemShut {NoStop}%
\bibitem [{\citenamefont {Laurell}\ \emph {et~al.}(2023)\citenamefont {Laurell}, \citenamefont {Luo}, \citenamefont {Weissenbilder}, \citenamefont {Ammitzböll}, \citenamefont {Ahmed}, \citenamefont {Söderberg}, \citenamefont {Petersson}, \citenamefont {Poulain}, \citenamefont {Guo}, \citenamefont {Dittel}, \citenamefont {Finkelstein-Shapiro}, \citenamefont {Squibb}, \citenamefont {Feifel}, \citenamefont {Gisselbrecht}, \citenamefont {Arnold}, \citenamefont {Buchleitner}, \citenamefont {Lindroth}, \citenamefont {Kockum}, \citenamefont {L'Huillier},\ and\ \citenamefont {Busto}}]{laurell_measuring_2023}%
  \BibitemOpen
  \bibfield  {author} {\bibinfo {author} {\bibfnamefont {H.}~\bibnamefont {Laurell}}, \bibinfo {author} {\bibfnamefont {S.}~\bibnamefont {Luo}}, \bibinfo {author} {\bibfnamefont {R.}~\bibnamefont {Weissenbilder}}, \bibinfo {author} {\bibfnamefont {M.}~\bibnamefont {Ammitzböll}}, \bibinfo {author} {\bibfnamefont {S.}~\bibnamefont {Ahmed}}, \bibinfo {author} {\bibfnamefont {H.}~\bibnamefont {Söderberg}}, \bibinfo {author} {\bibfnamefont {C.~L.~M.}\ \bibnamefont {Petersson}}, \bibinfo {author} {\bibfnamefont {V.}~\bibnamefont {Poulain}}, \bibinfo {author} {\bibfnamefont {C.}~\bibnamefont {Guo}}, \bibinfo {author} {\bibfnamefont {C.}~\bibnamefont {Dittel}}, \bibinfo {author} {\bibfnamefont {D.}~\bibnamefont {Finkelstein-Shapiro}}, \bibinfo {author} {\bibfnamefont {R.~J.}\ \bibnamefont {Squibb}}, \bibinfo {author} {\bibfnamefont {R.}~\bibnamefont {Feifel}}, \bibinfo {author} {\bibfnamefont {M.}~\bibnamefont {Gisselbrecht}}, \bibinfo {author} {\bibfnamefont {C.~L.}\ \bibnamefont {Arnold}}, \bibinfo {author}
  {\bibfnamefont {A.}~\bibnamefont {Buchleitner}}, \bibinfo {author} {\bibfnamefont {E.}~\bibnamefont {Lindroth}}, \bibinfo {author} {\bibfnamefont {A.~F.}\ \bibnamefont {Kockum}}, \bibinfo {author} {\bibfnamefont {A.}~\bibnamefont {L'Huillier}},\ and\ \bibinfo {author} {\bibfnamefont {D.}~\bibnamefont {Busto}},\ }\href {http://arxiv.org/abs/2309.13945} {\bibinfo {title} {Measuring the quantum state of photoelectrons}} (\bibinfo {year} {2023}),\ \bibinfo {note} {arXiv:2309.13945 [quant-ph]}\BibitemShut {NoStop}%
\bibitem [{\citenamefont {Kato}\ and\ \citenamefont {Kono}(2009)}]{kato_time-dependent_2009}%
  \BibitemOpen
  \bibfield  {author} {\bibinfo {author} {\bibfnamefont {T.}~\bibnamefont {Kato}}\ and\ \bibinfo {author} {\bibfnamefont {H.}~\bibnamefont {Kono}},\ }\bibfield  {title} {\bibinfo {title} {Time-dependent multiconfiguration theory for ultrafast electronic dynamics of molecules in an intense laser field: {Electron} correlation and energy redistribution among natural orbitals},\ }\href {https://doi.org/10.1016/j.chemphys.2009.09.017} {\bibfield  {journal} {\bibinfo  {journal} {Chemical Physics}\ }\bibinfo {series} {Attosecond {Molecular} {Dynamics}},\ \textbf {\bibinfo {volume} {366}},\ \bibinfo {pages} {46} (\bibinfo {year} {2009})}\BibitemShut {NoStop}%
\bibitem [{\citenamefont {Kramida}\ \emph {et~al.}(2024)\citenamefont {Kramida}, \citenamefont {{Yu.~Ralchenko}}, \citenamefont {Reader},\ and\ \citenamefont {{and NIST ASD Team}}}]{NIST_ASD}%
  \BibitemOpen
  \bibfield  {author} {\bibinfo {author} {\bibfnamefont {A.}~\bibnamefont {Kramida}}, \bibinfo {author} {\bibnamefont {{Yu.~Ralchenko}}}, \bibinfo {author} {\bibfnamefont {J.}~\bibnamefont {Reader}},\ and\ \bibinfo {author} {\bibnamefont {{and NIST ASD Team}}},\ }\href@noop {} {}\bibinfo {howpublished} {{NIST Atomic Spectra Database (ver. 5.12), [Online]. Available: {\tt{https://physics.nist.gov/asd}} [2024, December 23]. National Institute of Standards and Technology, Gaithersburg, MD.}} (\bibinfo {year} {2024})\BibitemShut {NoStop}%
\bibitem [{\citenamefont {Romanov}\ \emph {et~al.}(2021)\citenamefont {Romanov}, \citenamefont {Silaev}, \citenamefont {Sarantseva}, \citenamefont {Frolov},\ and\ \citenamefont {Vvedenskii}}]{romanov_study_2021}%
  \BibitemOpen
  \bibfield  {author} {\bibinfo {author} {\bibfnamefont {A.~A.}\ \bibnamefont {Romanov}}, \bibinfo {author} {\bibfnamefont {A.~A.}\ \bibnamefont {Silaev}}, \bibinfo {author} {\bibfnamefont {T.~S.}\ \bibnamefont {Sarantseva}}, \bibinfo {author} {\bibfnamefont {M.~V.}\ \bibnamefont {Frolov}},\ and\ \bibinfo {author} {\bibfnamefont {N.~V.}\ \bibnamefont {Vvedenskii}},\ }\bibfield  {title} {\bibinfo {title} {Study of high-order harmonic generation in xenon based on time-dependent density-functional theory},\ }\href {https://doi.org/10.1088/1367-2630/abe8a9} {\bibfield  {journal} {\bibinfo  {journal} {New Journal of Physics}\ }\textbf {\bibinfo {volume} {23}},\ \bibinfo {pages} {043014} (\bibinfo {year} {2021})}\BibitemShut {NoStop}%
\bibitem [{\citenamefont {Sato}\ and\ \citenamefont {Ishikawa}(2013)}]{sato_time-dependent_2013}%
  \BibitemOpen
  \bibfield  {author} {\bibinfo {author} {\bibfnamefont {T.}~\bibnamefont {Sato}}\ and\ \bibinfo {author} {\bibfnamefont {K.~L.}\ \bibnamefont {Ishikawa}},\ }\bibfield  {title} {\bibinfo {title} {Time-dependent complete-active-space self-consistent-field method for multielectron dynamics in intense laser fields},\ }\href {https://doi.org/10.1103/PhysRevA.88.023402} {\bibfield  {journal} {\bibinfo  {journal} {Physical Review A}\ }\textbf {\bibinfo {volume} {88}},\ \bibinfo {pages} {023402} (\bibinfo {year} {2013})}\BibitemShut {NoStop}%
\bibitem [{\citenamefont {Lackner}\ \emph {et~al.}(2015)\citenamefont {Lackner}, \citenamefont {Brezinova}, \citenamefont {Sato}, \citenamefont {Ishikawa},\ and\ \citenamefont {Burgdörfer}}]{lackner_propagating_2015}%
  \BibitemOpen
  \bibfield  {author} {\bibinfo {author} {\bibfnamefont {F.}~\bibnamefont {Lackner}}, \bibinfo {author} {\bibfnamefont {I.}~\bibnamefont {Brezinova}}, \bibinfo {author} {\bibfnamefont {T.}~\bibnamefont {Sato}}, \bibinfo {author} {\bibfnamefont {K.~L.}\ \bibnamefont {Ishikawa}},\ and\ \bibinfo {author} {\bibfnamefont {J.}~\bibnamefont {Burgdörfer}},\ }\bibfield  {title} {\bibinfo {title} {Propagating two-particle reduced density matrices without wavefunctions},\ }\href {https://doi.org/10.1103/PhysRevA.91.023412} {\bibfield  {journal} {\bibinfo  {journal} {Physical Review A}\ }\textbf {\bibinfo {volume} {91}},\ \bibinfo {pages} {023412} (\bibinfo {year} {2015})},\ \bibinfo {note} {arXiv:1411.0495 [physics]}\BibitemShut {NoStop}%
\bibitem [{\citenamefont {Lackner}\ \emph {et~al.}(2017)\citenamefont {Lackner}, \citenamefont {Březinová}, \citenamefont {Sato}, \citenamefont {Ishikawa},\ and\ \citenamefont {Burgdörfer}}]{lackner_high-harmonic_2017}%
  \BibitemOpen
  \bibfield  {author} {\bibinfo {author} {\bibfnamefont {F.}~\bibnamefont {Lackner}}, \bibinfo {author} {\bibfnamefont {I.}~\bibnamefont {Březinová}}, \bibinfo {author} {\bibfnamefont {T.}~\bibnamefont {Sato}}, \bibinfo {author} {\bibfnamefont {K.~L.}\ \bibnamefont {Ishikawa}},\ and\ \bibinfo {author} {\bibfnamefont {J.}~\bibnamefont {Burgdörfer}},\ }\bibfield  {title} {\bibinfo {title} {High-harmonic spectra from time-dependent two-particle reduced-density-matrix theory},\ }\href {https://doi.org/10.1103/PhysRevA.95.033414} {\bibfield  {journal} {\bibinfo  {journal} {Physical Review A}\ }\textbf {\bibinfo {volume} {95}},\ \bibinfo {pages} {033414} (\bibinfo {year} {2017})}\BibitemShut {NoStop}%
\bibitem [{\citenamefont {Donsa}\ \emph {et~al.}(2023)\citenamefont {Donsa}, \citenamefont {Lackner}, \citenamefont {Burgd\"orfer}, \citenamefont {Bonitz}, \citenamefont {Kloss}, \citenamefont {Rubio},\ and\ \citenamefont {B\ifmmode~\check{r}\else \v{r}\fi{}ezinov\'a}}]{donsa_nonequilibrium_2023}%
  \BibitemOpen
  \bibfield  {author} {\bibinfo {author} {\bibfnamefont {S.}~\bibnamefont {Donsa}}, \bibinfo {author} {\bibfnamefont {F.}~\bibnamefont {Lackner}}, \bibinfo {author} {\bibfnamefont {J.}~\bibnamefont {Burgd\"orfer}}, \bibinfo {author} {\bibfnamefont {M.}~\bibnamefont {Bonitz}}, \bibinfo {author} {\bibfnamefont {B.}~\bibnamefont {Kloss}}, \bibinfo {author} {\bibfnamefont {A.}~\bibnamefont {Rubio}},\ and\ \bibinfo {author} {\bibfnamefont {I.}~\bibnamefont {B\ifmmode~\check{r}\else \v{r}\fi{}ezinov\'a}},\ }\bibfield  {title} {\bibinfo {title} {Nonequilibrium correlation dynamics in the one-dimensional fermi-hubbard model: A testbed for the two-particle reduced density matrix theory},\ }\href {https://doi.org/10.1103/PhysRevResearch.5.033022} {\bibfield  {journal} {\bibinfo  {journal} {Phys. Rev. Res.}\ }\textbf {\bibinfo {volume} {5}},\ \bibinfo {pages} {033022} (\bibinfo {year} {2023})}\BibitemShut {NoStop}%
\bibitem [{\citenamefont {Pescoller}\ \emph {et~al.}(2025)\citenamefont {Pescoller}, \citenamefont {Eder},\ and\ \citenamefont {B\v{r}ezinov\'a}}]{Pescoller_projective_2025}%
  \BibitemOpen
  \bibfield  {author} {\bibinfo {author} {\bibfnamefont {E.}~\bibnamefont {Pescoller}}, \bibinfo {author} {\bibfnamefont {M.}~\bibnamefont {Eder}},\ and\ \bibinfo {author} {\bibfnamefont {I.}~\bibnamefont {B\v{r}ezinov\'a}},\ }\bibfield  {title} {\bibinfo {title} {Projective purification of correlated reduced density matrices},\ }\href {https://doi.org/10.1103/PhysRevResearch.7.013211} {\bibfield  {journal} {\bibinfo  {journal} {Phys. Rev. Res.}\ }\textbf {\bibinfo {volume} {7}},\ \bibinfo {pages} {013211} (\bibinfo {year} {2025})}\BibitemShut {NoStop}%
\bibitem [{\citenamefont {Rescigno}\ and\ \citenamefont {McCurdy}(2000)}]{rescigno_numerical_2000}%
  \BibitemOpen
  \bibfield  {author} {\bibinfo {author} {\bibfnamefont {T.~N.}\ \bibnamefont {Rescigno}}\ and\ \bibinfo {author} {\bibfnamefont {C.~W.}\ \bibnamefont {McCurdy}},\ }\bibfield  {title} {\bibinfo {title} {Numerical grid methods for quantum-mechanical scattering problems},\ }\href {https://doi.org/10.1103/PhysRevA.62.032706} {\bibfield  {journal} {\bibinfo  {journal} {Physical Review A}\ }\textbf {\bibinfo {volume} {62}},\ \bibinfo {pages} {032706} (\bibinfo {year} {2000})},\ \bibinfo {note} {publisher: American Physical Society}\BibitemShut {NoStop}%
\bibitem [{\citenamefont {Auzinger}\ \emph {et~al.}(2021)\citenamefont {Auzinger}, \citenamefont {Březinová}, \citenamefont {Grosz}, \citenamefont {Hofstätter}, \citenamefont {Koch},\ and\ \citenamefont {Sato}}]{auzinger_efficient_2021}%
  \BibitemOpen
  \bibfield  {author} {\bibinfo {author} {\bibfnamefont {W.}~\bibnamefont {Auzinger}}, \bibinfo {author} {\bibfnamefont {I.}~\bibnamefont {Březinová}}, \bibinfo {author} {\bibfnamefont {A.}~\bibnamefont {Grosz}}, \bibinfo {author} {\bibfnamefont {H.}~\bibnamefont {Hofstätter}}, \bibinfo {author} {\bibfnamefont {O.}~\bibnamefont {Koch}},\ and\ \bibinfo {author} {\bibfnamefont {T.}~\bibnamefont {Sato}},\ }\bibfield  {title} {\bibinfo {title} {Efficient adaptive exponential time integrators for nonlinear {Schrödinger} equations with nonlocal potential},\ }\href {https://doi.org/10.1016/j.jcmds.2021.100014} {\bibfield  {journal} {\bibinfo  {journal} {Journal of Computational Mathematics and Data Science}\ }\textbf {\bibinfo {volume} {1}},\ \bibinfo {pages} {100014} (\bibinfo {year} {2021})}\BibitemShut {NoStop}%
\bibitem [{\citenamefont {Lehtola}\ \emph {et~al.}(2018)\citenamefont {Lehtola}, \citenamefont {Steigemann}, \citenamefont {Oliveira},\ and\ \citenamefont {Marques}}]{lehtola_recent_2018}%
  \BibitemOpen
  \bibfield  {author} {\bibinfo {author} {\bibfnamefont {S.}~\bibnamefont {Lehtola}}, \bibinfo {author} {\bibfnamefont {C.}~\bibnamefont {Steigemann}}, \bibinfo {author} {\bibfnamefont {M.~J.}\ \bibnamefont {Oliveira}},\ and\ \bibinfo {author} {\bibfnamefont {M.~A.}\ \bibnamefont {Marques}},\ }\bibfield  {title} {\bibinfo {title} {Recent developments in libxc — {A} comprehensive library of functionals for density functional theory},\ }\href {https://doi.org/10.1016/j.softx.2017.11.002} {\bibfield  {journal} {\bibinfo  {journal} {SoftwareX}\ }\textbf {\bibinfo {volume} {7}},\ \bibinfo {pages} {1} (\bibinfo {year} {2018})}\BibitemShut {NoStop}%
\bibitem [{\citenamefont {Perdew}\ \emph {et~al.}(1997)\citenamefont {Perdew}, \citenamefont {Burke},\ and\ \citenamefont {Ernzerhof}}]{perdew_generalized_1997}%
  \BibitemOpen
  \bibfield  {author} {\bibinfo {author} {\bibfnamefont {J.~P.}\ \bibnamefont {Perdew}}, \bibinfo {author} {\bibfnamefont {K.}~\bibnamefont {Burke}},\ and\ \bibinfo {author} {\bibfnamefont {M.}~\bibnamefont {Ernzerhof}},\ }\bibfield  {title} {\bibinfo {title} {Generalized {Gradient} {Approximation} {Made} {Simple} [{Phys}. {Rev}. {Lett}. 77, 3865 (1996)]},\ }\href {https://doi.org/10.1103/physrevlett.78.1396} {\bibfield  {journal} {\bibinfo  {journal} {Physical Review Letters}\ }\textbf {\bibinfo {volume} {78}},\ \bibinfo {pages} {1396} (\bibinfo {year} {1997})},\ \bibinfo {note} {publisher: American Physical Society (APS)}\BibitemShut {NoStop}%
\bibitem [{\citenamefont {Gaiduk}\ and\ \citenamefont {Staroverov}(2011)}]{gaiduk_construction_2011}%
  \BibitemOpen
  \bibfield  {author} {\bibinfo {author} {\bibfnamefont {A.~P.}\ \bibnamefont {Gaiduk}}\ and\ \bibinfo {author} {\bibfnamefont {V.~N.}\ \bibnamefont {Staroverov}},\ }\bibfield  {title} {\bibinfo {title} {Construction of integrable model {Kohn}-{Sham} potentials by analysis of the structure of functional derivatives},\ }\bibfield  {journal} {\bibinfo  {journal} {Physical Review A}\ }\textbf {\bibinfo {volume} {83}},\ \href {https://doi.org/10.1103/physreva.83.012509} {10.1103/physreva.83.012509} (\bibinfo {year} {2011}),\ \bibinfo {note} {publisher: American Physical Society (APS)}\BibitemShut {NoStop}%
\bibitem [{\citenamefont {Vosko}\ \emph {et~al.}(1980)\citenamefont {Vosko}, \citenamefont {Wilk},\ and\ \citenamefont {Nusair}}]{vosko_accurate_1980}%
  \BibitemOpen
  \bibfield  {author} {\bibinfo {author} {\bibfnamefont {S.~H.}\ \bibnamefont {Vosko}}, \bibinfo {author} {\bibfnamefont {L.}~\bibnamefont {Wilk}},\ and\ \bibinfo {author} {\bibfnamefont {M.}~\bibnamefont {Nusair}},\ }\bibfield  {title} {\bibinfo {title} {Accurate spin-dependent electron liquid correlation energies for local spin density calculations: a critical analysis},\ }\href {https://doi.org/10.1139/p80-159} {\bibfield  {journal} {\bibinfo  {journal} {Canadian Journal of Physics}\ }\textbf {\bibinfo {volume} {58}},\ \bibinfo {pages} {1200} (\bibinfo {year} {1980})},\ \bibinfo {note} {publisher: Canadian Science Publishing}\BibitemShut {NoStop}%
\end{thebibliography}%

\end{document}